\documentclass[aps,amssymb,amsmath,prd,twocolumn,
showpacs,preprintnumbers,superscriptaddress,nofootinbib,floatfix]{revtex4-1}
\usepackage{graphicx}
\usepackage{bm}
\usepackage{mathrsfs}
\usepackage{latexsym}
\usepackage{float}
\usepackage{color}
\usepackage[normalem]{ulem} 
\usepackage{dcolumn}
\usepackage[colorlinks=true,citecolor=blue,urlcolor=blue]{hyperref}
\usepackage[usenames,dvipsnames]{xcolor}
\usepackage[caption=false]{subfig}
\usepackage{slashed}

\newcommand{\p}{\partial}
\newcommand{\nn}{\nonumber}
\newcommand{\be}{\begin{equation}}
\newcommand{\ee}{\end{equation}}
\newcommand{\beq}{\begin{eqnarray}}
\newcommand{\eeq}{\end{eqnarray}}

\newcommand{\nb}{n_{\rm B}}
\newcommand{\nsat}{n_{\rm sat}}
\newcommand{\Msolar}{{\rm M}_{\odot}}
\newcommand{\Mmax}{\ensuremath{M_{\rm max}}}

\newcommand{\ep}{\varepsilon}

\newcommand{\csqeq}{c_{\rm eq}^2}
\newcommand{\csqad}{c_{\rm ad}^2}

\newcommand{\bv}{Brunt-V\"ais\"al\"a}
\newcommand{\gm}{$g$-mode}

\begin{document}
\preprint{N3AS-23-004}
\title{Framework for phase transitions between the Maxwell and Gibbs constructions} 

\author{Constantinos Constantinou}
\email{cconstantinou@ectstar.eu}
\affiliation{INFN-TIFPA, Trento Institute of Fundamental Physics and Applications, Povo, 38123 TN, Italy}
\affiliation{European Centre for Theoretical Studies in Nuclear Physics and Related Areas, Villazzano, 38123 TN, Italy}

\author{Tianqi Zhao}
\email{zhaot@ohio.edu}
\affiliation{Department of Physics and Astronomy, Ohio University,
Athens, OH~45701, USA}

\author{Sophia Han}
\email{sjhan@sjtu.edu.cn}
\affiliation{Tsung-Dao Lee Institute, Shanghai Jiao Tong University, Shanghai 201210, China}
\affiliation{School of Physics and Astronomy, Shanghai Jiao Tong University, Shanghai 200240, China}

\author{Madappa Prakash}
\email{prakash@ohio.edu}
\affiliation{Department of Physics and Astronomy, Ohio University,
Athens, OH~45701, USA}

\date{\today}

\begin{abstract}

By taking the nucleon-to-quark phase transition within a neutron star as an example, we present a thermodynamically consistent method to calculate the equation of state of ambient matter so that transitions that are intermediate to those of the familiar Maxwell and Gibbs constructions can be described. This method does not address the poorly known surface tension between the two phases microscopically (as, for example, in the calculation of the core pasta phases via the Wigner-Seitz approximation) but instead combines the local and global charge neutrality conditions characteristic of the Maxwell and Gibbs constructions, respectively. Overall charge neutrality is achieved by dividing the leptons to those that obey local charge neutrality (Maxwell) and those that maintain global charge neutrality (Gibbs). The equation of state is obtained by using equilibrium constraints derived from minimizing the total energy density. The results of this minimization are then used to calculate neutron star mass-radius curves, tidal deformabilities, equilibrium and adiabatic sound speeds, and nonradial $g$-mode oscillation frequencies for several intermediate constructions. Various quantities of interest transform smoothly from their Gibbs structures to those of Maxwell as the local-to-total electron ratio $\eta$, introduced to mimic the hadron-to-quark interface tension from $0$ (Gibbs) to $\infty$ (Maxwell), is raised from $0$ to $1$. A notable exception is the $g$-mode frequency for the specific case of $\eta=1$ 
for which a gap appears between the quark and hadronic branches.

\end{abstract}

\maketitle


\section{Introduction}
\label{sec:intro}

In describing the transition from baryonic or hadronic matter to that of its constituents, such as up, down and strange quarks in the interiors of neutron stars (NSs), the most commonly employed methods are either the Maxwell or the Gibbs construction~\cite{Glendenning:1992vb,Glendenning:2001pe}. 
In the Maxwell construction, charge neutrality is achieved locally, whereas in the Gibbs construction the same is achieved globally. The Maxwell construction is applicable when the interface or surface tension between the two phases is very large,  
whereas the Gibbs construction is valid in the opposite limit of vanishing (or zero) surface tension. 
{(The phrases large and small here refer to whether or not the surface contribution to the Coulomb energy is large or small.)}
While the former method is suitable for transitions from a single component system (say, neutrons only), the latter is well suited when multiple charges such as neutrons, protons, and electrons are present in the transitioning system, particularly to account for separate baryon number conservation and charge neutrality. For intermediate surface tensions, the shape of the phase boundary could vary with density as in the pasta phase at the crust-core transition treated in the Wigner-Seitz approximation~\cite{Heiselberg:1992dx}. Nevertheless, the magnitude of the quark-hadron interface tension is highly uncertain, ranging from a few to hundreds of MeV/fm$^2$; see e.g. discussions in the literature~\cite{Alford:2001zr,Mintz:2009ay,Palhares:2010be,Lugones:2013ema,Lugones:2018qgu,Fraga:2018cvr,Schmitt:2020tac,Ju:2021hoy}. \\

In the pressure $P$ vs energy density $\ep$ plane, the Maxwell construction in which the pressure and neutron chemical potential equalities $P(H)=P(Q)$ and $\mu_n(H)=\mu_n(Q)$ are established between the hadronic ($H$) and quark ($Q$) phases is characterized by a flat region. The range of densities over which these equalities hold can be determined using the methods described in Refs.~\cite{Lamb:1983djd,Constantinou:2014hha}. 
A consequence of this flat region is that the squared equilibrium speed of sound $\csqeq=dP/d\ep$ becomes zero  there.
The density region over which the flat region occurs as well as the extent of the jump in the energy density depend on the details of the $P~ {\rm vs}~\ep$ relationships, or the equation of state (EOS), in each of the two phases. \\

The description of the mixed phase in the Gibbs construction is achieved by satisfying the rules $P(H)=P(Q)$ and $\mu_n(H)=\mu_u+2\mu_d$, where the chemical potentials $\mu_u$ and $\mu_d$ refer to those of the up ($u$) and down ($d$) quarks, respectively.  
The conditions of global charge neutrality and baryon number conservation are imposed through the relations 
\begin{eqnarray}
 Q &=& fQ(H) + (1-f)Q(Q) =0   \nonumber \\
 \nb &=& f\nb(H) + (1-f)\nb(Q) \,,   
\end{eqnarray}
where $f$ denotes the fractional volume occupied by hadrons and is solved for each baryon density $\nb$. Unlike in the pure phases of the Maxwell construction, $Q(H)$ and $Q(Q)$ do not separately vanish in the Gibbs mixed phase. The total energy density is given by
\begin{eqnarray}
\ep = f\ep(H) + (1-f)\ep(Q) \,.
\end{eqnarray}
Relative to the Maxwell construction, the behavior of the pressure vs baryon density is smooth in the case of Gibbs construction. 
Discontinuities in its derivatives with respect to baryon density, reflected in $\csqeq=dP/d\ep$, will however, be present at the densities where the mixed phase begins and ends. \\

Situations in which neither the Maxwell nor the Gibbs construction can be applied correspond to cases in which the pressure and chemical potential equalities cannot be met for many hadronic and quark EOSs. In such cases, interpolatory techniques that make the transition a smooth crossover have been used in Refs.~\cite{Baym:2017whm,Masuda:2012ed,Fukushima:2015bda,Kojo:2014rca}. 
In these approaches, the pressure equality between the two phases characteristic of Maxwell and Gibbs constructions is abandoned, but the pressure vs baryon density in the mixed phase is composed of contributions from hadrons and quarks in an externally prescribed proportion. Outside of the mixed phase, pure hadronic and quark phases exist. The onset and ending densities of the mixed phase are chosen suitably for smooth crossover from one phase to the other. \\

The model termed quarkyonic matter departs from first-order phase transitions inasmuch as once quarks appear, both nucleons and quarks coexist until asymptotically large baryon densities when the baryon concentrations vanish~\cite{McLerran:2018hbz}. 
The order of the phase transition depends on the implementation of these models. In Ref.~\cite{McLerran:2018hbz},  
the transition is second order, but other approaches~\cite{Masuda:2012ed,Kojo:2014rca,Fukushima:2015bda} 
have yielded higher-order phase transitions. A characteristic feature of the quarkyonic models is that the $\csqeq = dP/d\ep$ exhibits a peak before approaching the value of $1/3$, an attribute of asymptotically free quarks. Depending on the approach adopted, this value may also be reached from below~\cite{McLerran:2018hbz,Jeong:2019lhv,Sen:2020qcd}. A drawback of the quarkyonic model with a second-order transition is that the squared adiabatic speed of sound $\csqad = (\partial P/\partial \ep)_{y_p}$, where $y_p$ is the proton fraction, becomes infinite at the onset of quarks~\cite{Constantinou:2021hba}. This feature prevents the calculation of oscillation modes of NSs, particularly the $g$-modes (gravity modes)~\cite{Constantinou:2021hba}.  \\

A crossover model for the transition from hadrons to quarks in NSs to mimic the crossover feature of baryon-free finite temperature studies has also been investigated in Ref.~\cite{Kapusta:2021ney}. 
The key feature of this approach is an analytic mixing or switching function that accounts for the partial pressure of each component as a function of a single thermodynamic variable---the baryon chemical potential. As in the mixed phase hadrons/nucleons and quarks both appear  explicitly as separate degrees of freedom in this description, it is straightforward to keep track of their individual contributions to the total pressure. In Ref.~\cite{Constantinou:2021hba}, this approach was generalized to beta-equilibrated matter in order to explore nonradial $g$-mode oscillations of NSs.   \\

Our goal here is to devise a framework in which the Maxwell and Gibbs constructions are two extremes of a continuous spectrum of possibilities for first-order phase transitions. We accomplish our goal by postulating three distinct electron clouds (with labels $eN$, $eQ$, and $eG$), whose members are either strictly in contact with only nucleons ($eN$) or quarks ($eQ$), or can be shared between the two phases ($eG$). Thus, charge neutrality is fulfilled partially locally and partially globally, the ratio being controlled by a new variable $\eta$, 
which stands for the local-to-total electron ratio. Note that, here, we do not posit distinguishable electrons in the sense of intrinsic quantum numbers; instead, we simply group them in relation to the many-body environment in which they are embedded. This grouping is artificial and, at the end of the calculation, we will be interested only in the total number or fraction of electrons required for the system to be charge neutral. \\

The physical picture is as follows. In the case of a large surface tension between the hadron and the quark phases, the boundary between the two is sharp, and the region each phase occupies is well defined. Correspondingly, the electrons ensuring charge neutrality will be unequivocally associated with (or, at the very least, are far more likely to interact with) one or the other phase by virtue of their spatial position; thus charge neutrality is local. In the opposite limit of very-low/zero surface tension, there is no spatial separation between the two phases and thus charge neutrality is accomplished entirely globally. \\

For intermediate surface tension, the boundary between the two phases becomes fuzzy and therefore, in addition to the two unambiguous regions from before, we have a third, gray-zone region where the phase of baryonic matter is unclear. Consequently, some electrons will be explicitly attached to one or the other phase, while the rest interact with both. In this case, charge neutrality is fulfilled partially locally (by some electrons) and partially globally (by the remaining electrons).  \\

{The precise mapping of the surface tension to the variable $\eta$ would require a specific model (that we have not considered) for the surface tension. As many models for the surface tension with hugely varying values exist in the literature (see Refs. \cite{Alford:2001zr,Mintz:2009ay,Palhares:2010be,Lugones:2013ema,Lugones:2018qgu,Fraga:2018cvr,Schmitt:2020tac,Ju:2021hoy}), such a mapping would differ from case to case depending on the model considered.  Even then, an $\eta$ that varies with density might be required. } \\

{Our framework offers a different way of modeling the mixed phase 
between the Maxwell (corresponding to $\eta=1$ equivalent to a large surface tension) and Gibbs (with $\eta=0$ equivalent to a small surface tension) constructions, so the extreme cases have a precise correspondence. Intermediate values $0<\eta<1$ would then represent small to large values of the surface tension, the precise one-to-one correspondence between $\eta$ and  the surface tension necessarily depending on the model chosen for the latter.   It is, however, a useful framework to provide EOSs as well as their particle compositions in the mixture for phase transitions between the Maxwell and Gibbs constructions. } \\

The organization of this paper is as follows. In Sec. II, the formalism to obtain a continuous spectrum of possibilities between the Maxwell and Gibbs constructions is detailed. Here, the relevant equations to describe matter with nucleons, quarks, and electrons as well as those including muons are provided. The equations of state for nucleons, quarks, leptons, and the squared equilibrium and adiabatic sound speeds are given in Sec. III. Non-radial $g$-mode oscillations are discussed in Sec. IV.  Results of our calculations are presented in Sec. V. A summary and conclusions are contained in Sec. VI.    \\

\section{Simulating transitions between Maxwell and Gibbs constructions}
\label{sec:framework}

In this section, we present the formalism to obtain a continuous spectrum of possibilities between the Maxwell and Gibbs constructions for first-order phase transitions. We begin with matter containing neutrons and protons, or nucleons ($N$), quarks ($Q$) and electrons ($e$) only. Thereafter, the discussion includes muons ($\mu$) as well. Relations corresponding to the conservation laws of baryon number and charge neutrality that connect the various particle fractions $y_i$, with $i$ covering $N=n,p$, $Q=u,d,s$, and the volume fractions $f$ and $\eta$ are presented first. The working equations result from energy density minimization with respect to the list of variables in $NQe\mu$ and $f$. Values of the local-to-total electron ratio $\eta$ are chosen parametrically in the range $(0,1)$. \\

\subsection{$NQe$ matter}
The total energy density of the system is given by the sum of appropriately weighted contributions from the individual components,
\begin{eqnarray}
  \ep &=& f(\ep_n + \ep_p + \eta\ep_{eN}) \nonumber \\
            &+& (1 - f)(\ep_u + \ep_d + \ep_s + \eta\ep_{eQ}) \nonumber \\
            &+&(1 - \eta)\ep_{eG} \,, \label{edenmix}  
\end{eqnarray}
where $f$ is the hadron-to-baryon fraction and $\eta$ is the ratio of electrons participating in local charge neutrality to the total number of electrons. 

Baryon and lepton conservation correspond to the equations
\begin{eqnarray}
1 &=& f(y_n + y_p) + (1 -f)(y_u + y_d + y_s)/3 \label{baryonnum} \\
0 &=& y_e - f\eta y_{eN} - (1-f)\eta y_{eQ} - (1 - \eta)y_{eG} \label{electronnum} ~,
\end{eqnarray}
whereas charge neutrality is described by the relations
\begin{eqnarray}   
0 &=& (y_p - y_{eN})\eta   \label{lcnH} \\ 
0 &=& [(2y_u - y_d - y_s)/3 - y_{eQ}]\eta \label{lcnQ} \\
0 &=& [fy_p + (1 - f)(2y_u - y_d - y_s)/3 - y_{eG}](1-\eta) \label{gcn} ~.        
\end{eqnarray}
The overall factors of $\eta$ and $(1-\eta)$ in Eqs.~(\ref{lcnH})-(\ref{gcn}), are not necessary but they have been kept to emphasize the fact that these equations describe \textit{partial} local charge neutrality (LCN) and global charge neutrality (GCN). 

Equations~(\ref{baryonnum})-(\ref{gcn}) are then used to eliminate 5 of the 12 free variables ($\nb$, $y_n$, $y_p$, $y_u$, $y_d$, $y_s$, $y_{eN}$, $y_{eQ}$, $y_{eG}$, $y_e$, $f$, $\eta$) 
in this scheme. The choice is arbitrary but the most convenient set (that is, the set that leads to physically transparent phase-equilibrium conditions in the fewest number of operations) is the following:
\begin{eqnarray}
y_u &=& \frac{1 + y_e - fy_n - 2fy_p}{1-f}  \\
y_d &=& \frac{2 - y_e - 2fy_n - fy_p - y_s(1-f)}{1-f}  \\
y_{eN} &=& y_p  \\
y_{eQ} &=& \frac{y_e - fy_p}{1-f}  \\
y_{eG} &=& y_e 
\end{eqnarray}
For the subsequent calculation, the nonzero partial derivatives of the above are necessary:
\begin{eqnarray}
\frac{\partial y_u}{\partial y_n} &=& \frac{-f}{1 - f}~,~~~ 
\frac{\partial y_u}{\partial y_p} = \frac{-2f}{1 - f}~,~~~
\frac{\partial y_u}{\partial y_e} = \frac{1}{1 - f}~,~~~ \nonumber \\
\frac{\partial y_u}{\partial f} &=& \frac{y_u - y_n - 2y_p}{1 - f}  \\
\frac{\partial y_d}{\partial y_n} &=& \frac{-2f}{1 - f}~,~~~ 
\frac{\partial y_d}{\partial y_p} = \frac{-f}{1 - f}~,~~~
\frac{\partial y_d}{\partial y_e} = \frac{-1}{1 - f}~,~~~   \nonumber \\
\frac{\partial y_d}{\partial y_s} &=& -1~,~~~
\frac{\partial y_d}{\partial f} = \frac{y_d + y_s - 2y_n - y_p}{1 - f}  \\
\frac{\partial y_{eN}}{\partial y_p} &=& 1   \\ 
\frac{\partial y_{eQ}}{\partial y_p} &=& \frac{-f}{1 - f}~,~~~ 
\frac{\partial y_{eQ}}{\partial y_e} = \frac{1}{1 - f}~,~~~ \nonumber \\
\frac{\partial y_{eQ}}{\partial y_e} &=& \frac{y_{eQ}-y_p}{1 - f}~,~~~  \\
\frac{\partial y_{eG}}{\partial y_e} &=& 1  
\end{eqnarray}
The ground state of matter is obtained by minimizing the energy density $\ep$ with respect to the remaining free variables [except the baryon density $\nb$ being that we want to retain it as a free variable for the purposes of studying neutron-star matter (NSM)]: \\
(a) The usual condition for neutron strong equilibrium results from minimization with respect to the neutron fraction, $y_n$.
\begin{eqnarray}
\frac{\partial \ep}{\partial y_n} &=& f \frac{\partial \ep_n}{\partial y_n}
+ (1-f)\left(\frac{\partial y_u}{\partial y_n}\frac{\partial \ep_u}{\partial y_u}
             + \frac{\partial y_d}{\partial y_n}\frac{\partial \ep_d}{\partial y_d}\right) 
             \nonumber \\
&=& f\nb\mu_n + (1 - f)\left(\frac{-f}{1-f}\nb\mu_u - \frac{2f}{1-f}\nb\mu_d\right)   \nonumber \\
&=& f\nb(\mu_n - \mu_u - 2\mu_d) = 0  \nonumber \\
\Rightarrow ~ \mu_n &=& \mu_u + 2\mu_d    \label{muneq}
\end{eqnarray}
(b) Minimization with respect to the proton fraction $y_p$ leads to a condition that combines proton strong and electron electromagnetic equilibrium. These two are no longer independent as a result of our having overspecified the system.
\begin{eqnarray}
\frac{\partial \ep}{\partial y_p} &=& 
f \left(\frac{\partial \ep_p}{\partial y_p} 
+ \eta\frac{\partial y_{eN}}{\partial y_p}\frac{\partial \ep_{eN}}{\partial y_{eN}} \right)  \nonumber \\
&+& (1-f)\left(\frac{\partial y_u}{\partial y_p}\frac{\partial \ep_u}{\partial y_u}
             + \frac{\partial y_d}{\partial y_p}\frac{\partial \ep_d}{\partial y_d}
             + \eta\frac{\partial y_{eQ}}{\partial y_p}\frac{\partial \ep_{eQ}}{\partial y_{eQ}}\right) 
             \nonumber \\
&=& f(\nb\mu_p + \eta\nb\mu_{eN})+(1-f) \nonumber \\
&\times& \left(\frac{-2f}{1-f}\nb\mu_u - \frac{f}{1-f}\nb\mu_d - \eta\frac{f}{1-f}\nb\mu_{eQ}\right)   \nonumber \\
&=& f\nb(\mu_p + \eta\mu_{eN} - 2\mu_u - \mu_d - \eta\mu_{eQ}) = 0  \nonumber \\
\Rightarrow ~ \mu_p &=& 2\mu_u + \mu_d - \eta(\mu_{eN} - \mu_{eQ}) \label{mupeq}
\end{eqnarray}
By combining Eqs.~(\ref{muneq}) and (\ref{mupeq}), we find 
\begin{eqnarray}
\mu_u &=& 1/3(2\mu_p - \mu_n + 2\Delta_{\eta})  \label{muuDelta}\\
\mu_d &=& 1/3(2\mu_n - \mu_p - \Delta_{\eta})  \label{mudDelta}\\
\Delta_{\eta} &\equiv& \eta(\mu_{eN} - \mu_{eQ})~. \label{Deltag}
\end{eqnarray}
(c) A chemical potential relation corresponding to quark $\beta$-equilibrium is obtained by minimizing with respect to the total electron fraction $y_e$,
\begin{eqnarray}
\frac{\partial \ep}{\partial y_e} &=& 
(1-f)\left(\frac{\partial y_u}{\partial y_e}\frac{\partial \ep_u}{\partial y_u} 
           + \frac{\partial y_d}{\partial y_e}\frac{\partial \ep_d}{\partial y_d}
           + \eta\frac{\partial y_{eQ}}{\partial y_e}\frac{\partial \ep_{eQ}}{\partial y_{eQ}}\right)  \nonumber \\
           &+& (1 - \eta)\frac{\partial y_{eG}}{\partial y_e}\frac{\partial \ep_{eG}}{\partial y_{eG}}
             \nonumber \\
&=& (1 - f) \nonumber \\
&\times& \left(\frac{1}{1-f}\nb\mu_u - \frac{1}{1-f}\nb\mu_d - \eta\frac{1}{1-f}\nb\mu_{eQ}\right) \nonumber \\
     &+& (1 - \eta)(1)\nb\mu_{eG}   \nonumber \\
&=& \nb[\mu_u - \mu_d + \eta\mu_{eQ} + (1 - \eta)\mu_{eG}] = 0  \nonumber \\
\Rightarrow ~ \mu_d &=& \mu_u + \eta\mu_{eQ} + (1 - \eta)\mu_{eG} \,.\label{betaQ}
\end{eqnarray}
This, together with Eq.~(\ref{mupeq}) engenders a relation for nucleon $\beta$-equilibrium,
\begin{eqnarray}
\mu_p &=& 2\mu_u + \mu_d - \eta\mu_{eN} +[\mu_d - \mu_u - (1 - \eta)\mu_{eG}]\nonumber \\
    &=& (\mu_u + 2\mu_d) - \eta\mu_{eN} - (1 - \eta)\mu_{eG}\nonumber \\
\Rightarrow ~ \mu_p &=& \mu_n - \eta\mu_{eN} - (1 - \eta)\mu_{eG}  \label{betaN}
\end{eqnarray}
(d) Minimization with respect to the strange-quark fraction $y_s$ gives a condition for quark weak equilibrium,
\begin{eqnarray}
\frac{\partial \ep}{\partial y_s} &=& 
(1-f)\left(\frac{\partial y_d}{\partial y_s}\frac{\partial \ep_d}{\partial y_d}
           + \frac{\partial \ep_s}{\partial y_s}\right) 
             \nonumber \\
&=& (1 - f)(-\nb\mu_d + \nb\mu_s) = 0   \nonumber \\
\Rightarrow ~ \mu_d &=& \mu_s  \label{weakQ}
\end{eqnarray}
This condition is necessary not only in neutron-star matter but also for supernovae and NS mergers where the relevant dynamical timescales are longer than those of quark flavor-changing processes.   \\
(e) We get the condition for mechanical equilibrium by minimizing the energy density with respect to $f$,
\begin{eqnarray}
\frac{\partial \ep}{\partial f} &=& 
(\ep_n + \ep_p + \eta\ep_{eN})
- (\ep_u + \ep_d + \ep_s + \eta\ep_{eQ})  \nonumber \\
&+& (1-f)\left(\frac{\partial y_u}{\partial f}\frac{\partial \ep_u}{\partial y_u}
              + \frac{\partial y_d}{\partial f}\frac{\partial \ep_d}{\partial y_d}
              + \eta\frac{\partial y_{eQ}}{\partial f}\frac{\partial \ep_{eQ}}{\partial y_{eQ}}\right) 
              \nonumber \\
&=& \ep_N + \eta\ep_{eN} - \ep_Q - \eta\ep_{eQ}  \nonumber \\
&+&  (1 - f)\left(\frac{y_u - y_n - 2y_p}{1 - f}\nb\mu_u \right. \nonumber \\
                &+& \left. \frac{y_d + y_s - 2y_n - y_p}{1 - f}\nb\mu_d 
                + \eta\frac{y_{eQ} - y_p}{1 - f}\nb\mu_{eQ}\right)  \nonumber \\
\end{eqnarray}
where, in going from the first to the second equality, use of $\partial \ep/\partial y_i = \nb \mu_i$ was made, together with the definitions $\ep_N \equiv \Sigma_{h=n,p} \,\ep_h$ and $\ep_Q \equiv \Sigma_{q=u,d,s}\,\ep_q$. In the next step, we group chemical potentials according to whether they are multiplied by nucleon or quark particle fractions,
\begin{eqnarray}
\frac{\partial \ep}{\partial f} 
&=& \ep_N + \eta\ep_{eN} \nonumber \\
&+& [(-\ep_Q +\nb y_u\mu_u + \nb y_d\mu_d + \nb y_s\mu_d) \nonumber \\
&+& \eta(-\ep_{eQ} + \nb y_{eQ}\mu_{eQ})] 
\nonumber \\
&-& (y_n + 2y_p)\nb\mu_u - (2y_n + y_p)\nb\mu_d - \eta \nb y_p\mu_{eQ} \nonumber \\
\end{eqnarray}
Then, those $\mu_u$ and $\mu_d$ that are proportional to $y_n$ and $y_p$ are replaced by Eqs. (\ref{muuDelta}) and (\ref{mudDelta}). Moreover, the term $\nb y_s\mu_d$ becomes $\nb y_s\mu_s$ [using Eq.~(\ref{weakQ})] with the whole parenthesis in which it belongs written as $P_Q$ (as per the $T=0$ thermodynamic identity $P=\nb\mu -\ep$),
\begin{eqnarray} 
\frac{\partial \ep}{\partial f}  
&=& \ep_N + \eta\ep_{eN} + P_Q + \eta P_{Qe}  \nonumber \\
&-& (y_n + 2y_p)\frac{\nb}{3}(2\mu_p - \mu_n + 2\Delta_{\eta}) \nonumber  \\
&-& (2y_n + y_p)\frac{\nb}{3}(2\mu_n - \mu_p - \Delta_{\eta}) - \eta \nb y_p\mu_{eQ} 
\nonumber \\
\end{eqnarray}
Subsequently, we expand the products in the second and third lines above and collect similar terms,
\begin{eqnarray}
\frac{\partial \ep}{\partial f} 
&=& \ep_N + \eta\ep_{eN} + P_Q + \eta P_{Qe} - \eta \nb y_p\mu_{eQ} \nonumber \\
&-& \frac{\nb}{3} (2y_n\mu_p - y_n\mu_n + 2\Delta_{\eta} y_n + 4y_p - 2y_p\mu_n + 4y_p\Delta_{\eta} \nonumber \\
&-&2y_n\mu_p + 4y_n\mu_n - 2y_n\Delta_{\eta} - y_p\mu_p + 2y_p\mu_n - y_p\Delta_{\eta})  \nonumber \\
&=& \ep_N + \eta\ep_{eN} + P_Q + \eta P_{Qe} - \eta \nb y_p\mu_{eQ} \nonumber \\
&-& \frac{\nb}{3}(3y_n\mu_n + 3y_p\mu_p + 3y_p\Delta_{\eta}) 
\end{eqnarray} 
Finally, we apply Eq.~(\ref{Deltag}) to replace $\Delta_{\eta}$ with the electronic chemical potentials $\mu_{eN}$ and $\mu_{eQ}$, which leads to an expression involving only the pressures of the various components (using $P=\nb\mu - \ep$ where necessary),
\begin{eqnarray}
\frac{\partial \ep}{\partial f} 
&=& (\ep_N - \nb y_n\mu_n - \nb y_p\mu_p) + P_Q + \eta P_{Qe} \nonumber \\
&-& \nb y_p(\Delta_{\eta} + \eta\mu_{eQ}) \nonumber \\
&=& -P_N + P_Q + \eta P_{Qe} +\ep_{eN} - \eta \nb y_{eN}\mu_{eN} \nonumber \\
&=& - P_N - \eta P_{eN} + P_Q + \eta P_{eQ} = 0 
\nonumber \\
&\Rightarrow& P_N + \eta P_{eN} = P_Q + \eta P_{eQ}   \label{mecheq}
\end{eqnarray}
(f) For completeness, we also include the result of the minimization with respect to $\eta$. However, we will not be implementing this condition because we want $\eta$ to remain a free variable (along with $\nb$) in order to explore the effects of the changing surface tension,
\begin{eqnarray}
\frac{\partial \ep}{\partial \eta} &=& 
            f\ep_{eN} + (1 - f)\ep_{eQ} - \ep_{eG} = 0 \nonumber  \\
\Rightarrow ~ \ep_{eG} &=& f\ep_{eN} + (1 - f)\ep_{eQ}  
\label{surface}
\end{eqnarray}
In the present approach, $\eta=0$ amounts to a Gibbs construction (GCN) and $\eta=1$ to a Maxwell construction (LCN). \textit{It has the added benefit of maintaining control over the various particle fractions in the Maxwell mixed phase, which has not been the case in previous literature.} Clearly, first-order transitions of intermediate surface tension will have $0 < \eta < 1$. Extension to finite temperature is accomplished by minimizing the free energy density instead of the energy density. The conservation laws remain the same, as do the formal expressions describing the phase-equilibrium conditions, albeit with the use of the corresponding finite-$T$ pressures and chemical potentials. Applications to supernovae and neutron star mergers require $(\nb,y_e,T)$ as independent variables; that is, one must also skip minimization with respect to $y_e$.  \\

\noindent\textbf{Crossovers.}---These can also be studied in this context. One sets $\eta=0$\footnote{Unlike first-order transitions where two distinct phases are in contact, crossovers involve only a single phase whose ground state properties change drastically as some parameter of the system is changed. Therefore, in the present context, electrons will always encounter a mixture of quarks and hadrons regardless of their configuration-space coordinates, and, correspondingly, charge neutrality is achieved globally, i.e., $\eta=0$.} and eliminates the mechanical equilibrium condition [Eq.~(\ref{mecheq})] in favor of an explicit functional form for $f$, which approaches asymptotically 0 and 1 at high and low densities, respectively, e.g., $f=1-\exp[-a~(\nb/\nsat)^{-b}]$, where $a$ and $b$ are fit parameters and $\nsat$ is the saturation density of symmetric nuclear matter. The hadron-to-baryon fraction $f$ can also depend on composition (prior to equilibration) with the added algebraic burden of terms proportional to $\partial f/\partial y_i$ in the equilibrium equations. 
\subsection{$NQe\mu$ matter}
The inclusion of muons in the calculation comes at the cost of four additional variables ($y_{\mu N}$, $y_{\mu Q}$, $y_{\mu G}$, $y_{\mu}$), a muon-number conservation equation that mimics Eq. (\ref{electronnum}) for electrons, and modifications to the total energy density of the system and the charge neutrality equations (baryon number and electron number equations are unaffected),
\begin{eqnarray}
\ep &=& f[\ep_n + \ep_p + \eta(\ep_{eN}+\ep_{\mu N})] \nonumber \\
            &+& (1 - f)[\ep_u + \ep_d + \ep_s + \eta(\ep_{eQ}+\ep_{\mu Q})] \nonumber \\
            &+&(1 - \eta)(\ep_{eG}+\ep_{\mu G}) \label{edenmixmu}\\ 
0 &=& (y_p - y_{eN}-y_{\mu N})\eta   \label{lcnHmu} \\ 
0 &=& [(2y_u - y_d - y_s)/3 - y_{eQ}-y_{\mu Q}]\eta \label{lcnQmu} \\
0 &=& [fy_p + (1 - f)(2y_u - y_d - y_s)/3 \nonumber \\
&-& y_{eG}-y_{\mu G}](1-\eta) \label{gcnmu}\\ 
0 &=& y_{\mu} - f\eta y_{\mu N} - (1-f)\eta y_{\mu Q} - (1 - \eta)y_{\mu G} \label{muonnum}  
\end{eqnarray}
The minimization procedure yields modifications to the mechanical equilibrium and surface-tension optimization conditions [Eqs. (\ref{mecheq}) and (\ref{surface})] such that muonic contributions are accounted, while the chemical potential relations [Eqs.~(\ref{muneq}),~(\ref{mupeq}),~(\ref{betaQ}) or (\ref{betaN}),~(\ref{weakQ})] remain unchanged. Moreover, three new constraints are generated corresponding to lepton weak equilibrium in each of the three regions,
\begin{eqnarray}
&&P_N + \eta(P_{eN} + P_{\mu N}) = P_Q + \eta(P_{eQ} + P_{\mu Q}) \\
&&\ep_{eG} + \ep_{\mu G} = f(\ep_{eN} + \ep_{\mu N}) + (1 - f)(\ep_{eQ} + \ep_{\mu Q}) \nonumber \\ \\
&&\mu_{eN} = \mu_{\mu N}~;~~\mu_{eQ} = \mu_{\mu Q}~;~~\mu_{eG} = \mu_{\mu G} 
\end{eqnarray}
%

\section{Equation of state}

To demonstrate the workings of the scheme devised above, we describe the EOSs employed for nucleons, quarks, and leptons below. Selected properties of NSs such as their mass-radius curves, equilibrium and adiabatic squared speeds of sound are calculated results of which are shown and discussed. The outer crust EOS described by a uniform background of relativistic degenerate electrons in an ionic lattice is relatively well understood. Here we use the SLy4 crust EOS for $\nb<0.05$ fm$^{-3}$~\cite{Chabanat:1997un,Douchin:2001sv}. 
As our focus is on the core $g$-modes, the composition information of the crust is ignored in calculating the equilibrium and adiabatic sound speeds (that is, the two speeds are set equal to each other).

\subsection{Nucleons}
For the description of nucleons, we use the Zhao-Lattimer (ZL) EOS ~\cite{Zhao:2020dvu} with the parametrization termed as ZLA in Ref.~\cite{Constantinou:2021hba}. The parameters of ZLA are detailed in Table 1. This is consistent with laboratory data at nuclear saturation density $\nsat\simeq 0.16~{\rm fm}^{-3}$, the chiral effective field theory calculations of Refs.~\cite{Drischler:2020hwi,Drischler:2020fvz}, and constraints obtained by Legred et al.~\cite{Legred:2021hdx}, which combined available observations including the radio pulsar mass measurements of PSR J0348+0432 and J0470+6620~\cite{Fonseca:2021wxt,Cromartie:2019kug,Antoniadis:2013pzd}, the mass and tidal deformability measurements of GW170817 and GW190425~\cite{Abbott:2018wiz,LIGO:2017qsa,Abbott:2020uma}, and the x-ray mass and radius constraints from latest \textit{NICER} measurements of J0030+0451 and J0470+6620~\cite{Miller:2019cac,Riley:2019yda,Miller:2021qha,Riley:2021pdl,Salmi:2022cgy}. The total energy density of nucleons with a common mass $m_N=939.5$ MeV is given by the density functional 
\begin{eqnarray}
\ep_N &=& \ep_N(\nb,y_n,y_p) \nn \\
&=& \frac{1}{8\pi^2\hbar^3}\sum_{h=n,p}\left\{k_{Fh}(k_{Fh}^2 + m_N^2)^{1/2}(2k_{Fh}^2 + m_N^2)\right.  \nn \\
 &-& \left. m_N^4\ln\left[\frac{k_{Fh}+(k_{Fh}^2 + m_N^2)^{1/2}}{m_N}\right]\right\} \nn \\
           &+& 4 \nb^2 y_n y_p \left\{\frac{a_0}{\nsat} 
            +\frac{b_0}{\nsat^{\gamma}} [\nb(y_n + y_p)]^{\gamma - 1}\right\}  \nn \\
            &+& \nb^2 (y_n - y_p)^2\left\{\frac{a_1}{\nsat} 
            + \frac{b_1}{\nsat^{\gamma_1}}[\nb(y_n + y_p)]^{\gamma_1-1}\right\}\, ,  
           \nn \\ \label{edenZL} 
\end{eqnarray}
where $k_{Fh} = (3\pi^2\hbar^3\nb y_h)^{1/3}$ is the Fermi momentum of nucleon species $h$. {Above and below units of $c=1$ are used; also, wherever $\hbar$ appears, $\hbar c$ is implied. }
The chemical potentials and the pressure are obtained from Eq.~(\ref{edenZL}) according to
\begin{eqnarray}
\mu_h &=& \frac{\partial (\ep_N/\nb)}{\partial y_h}  ~~;~~ h=n,p\\
P_N &=& \nb\sum_{h=n,p}\mu_h y_h - \ep_N ~.
\end{eqnarray}
\begin{table}[h]
\caption{Parameter sets used in the present work. {Units of $c=1$ are employed.}} 
\begin{center} 
\begin{tabular}{ccrc}
\hline
\hline
Model       & Parameter  & Value       & Units \\ \hline
            &  $a_0$     & -96.64    & MeV     \\
            &  $b_0$     &  58.85    & MeV     \\
  ZLA        &  $\gamma$  &  1.40     &         \\
            &  $a_1$     & -26.06    & MeV     \\
            &  $b_1$     &  7.34     & MeV     \\
            & $\gamma_1$ &  2.45     &         \\
\hline 
            & $m_u$      &  5.0     & MeV     \\
            & $m_d$      &  7.0     & MeV     \\
  vMIT      & $m_s$      &  150.0   & MeV     \\
            & $a$        & 0.20     & fm$^2$  \\
            & $B^{1/4}$  & 165.0    & MeV     \\
\hline
            & {$\hbar(c)$}    & 197.3     & {MeV~fm}    \\
 Constants  & $m_e$        & 0.511     & {MeV} \\
            & $m_{\mu}$    & 105.7     & {MeV} \\
\hline \hline
\end{tabular}
\end{center}
\label{tab:Parameters}
\end{table}
\subsection{Quarks}
For the calculation of the quark EOS, we use the vMIT bag model~\cite{Gomes:2018eiv, Klahn:2015mfa}. The total energy density of quarks in this context is 
\begin{eqnarray}
\ep_Q &=& \ep_Q(\nb,y_u,y_d,y_s) \nn \\ 
&=& \sum_{q=u,d,s}\ep_q + \frac{1}{2}a~\hbar~[\nb(y_u+y_d+y_s)]^2 
                + \frac{B}{\hbar^3}  \nn \\  \\
\ep_q &=& \frac{3}{8\pi^2\hbar^3}\left\{k_{Fq} (k_{Fq}^2 + m_q^2)^{1/2}(2 k_{Fq}^2 + m_q^2) 
\right.      \nn \\      
        &-& \left. m_q^4\ln\left[\frac{k_{Fq}+(k_{Fq}^2 + m_q^2)^{1/2}}{m_q}\right]\right\}  ~,
\end{eqnarray}
where $k_{Fq} = (\pi^2\hbar^3\nb y_q)^{1/3}$ is the Fermi momentum of quark species $q$. Similar to the nucleonic case, the chemical potentials and pressure can be derived from the thermodynamic identities
\begin{eqnarray}
\mu_q &=& \frac{\partial (\ep_Q/\nb)}{\partial y_q}  ~~;~~ q=u,d,s\\
P_Q &=& \nb\sum_{q=u,d,s}\mu_q y_q - \ep_Q ~.
\end{eqnarray}
The parameters of this EOS ($a$ amd $B$) referred to as vMIT in Table 1 are as shown there.

\subsection{Leptons}
Leptons are treated as noninteracting, relativistic particles for which
\begin{eqnarray}
\ep_L &=&\frac{1}{8\pi^2\hbar^3}\sum_l\left\{k_{Fl}(k_{Fl}^2+m_l^2)^{1/2}(2k_{Fl}^2 + m_l^2)\right.  \nn \\
      &-& \left. m_l^4\ln\left[\frac{k_{Fl}+(k_{Fl}^2 + m_l^2)^{1/2}}{m_l}\right]\right\}  \\
\mu_l &=& (k_{Fl}^2 + m_l^2)^{1/2} \\
P_L &=& \nb \sum_l y_l \mu_l - \ep_L \\
k_{Fl} &=& (3\pi^2\hbar^3 \nb y_l)^{1/3};~~ l=e,\mu~. 
\end{eqnarray}
At low baryon densities, only electrons are present in the system. The muon onset density is such that $\mu_e -m_{\mu} = 0$. Depending on the parametrization choice, this condition also gives the density at which muons vanish.

\subsection{Sound speeds in the pure and mixed phases}
We begin with pure-phase thermodynamic quantities written as functions of the total baryon density $\nb$, and the individual particle fractions $y_n$,~$y_p$,~$y_{eN}$,~$y_u$,~$y_d$,~$y_s$,~$y_{eQ}$,~$y_{eG}$,
\begin{eqnarray}
\ep_N &=& \ep_N(\nb,y_n,y_p) ~;~ P_N = P_N(\nb,y_n,y_p) ~; \nn \\
\mu_h &=& \mu_h(\nb,y_n,y_p) \label{pureN}\\ \nn \\  
\ep_Q &=& \ep_Q(\nb,y_u,y_d,y_s) ~;~ P_Q = P_Q(\nb,y_u,y_d,y_s) ~; \nn \\
\mu_q &=& \mu_q(\nb,y_q)  ~;~ q = u, d, s \label{pureQ} \\ \nn \\
\ep_{eX} &=& \ep_{eX}(\nb,y_{eX}) ~;~ P_{eX} = P_{eX}(\nb,y_{eX}) ~; \nn \\
\mu_{eX} &=& \mu_{eX}(\nb,y_{eX})  ~;~ X = N,Q,G\,.
\end{eqnarray}
In terms of these, we express the thermodynamics of the mixed 
$(^*)$ 
phase as
\begin{eqnarray}
\ep^* &=& f \ep_N + (1-f) \ep_Q \nn \\
&+& f\eta\ep_{eN} + (1-f)\eta\ep_{eQ} + (1-\eta)\ep_{eG}  \\
P^* &=&  f P_N + (1-f) P_Q \nn \\
&+& f\eta P_{eN} + (1-f)\eta P_{eQ} + (1-\eta)P_{eG}  \\
\mu_h^* &=& \mu_h ~~;~~ \mu_q^* = \mu_q   \\
y_h^* &=& f y_h  ~~;~~ y_q^* = (1-f) y_q \,.
\label{qnoG}
\end{eqnarray}
For NSM (denoted by the subscript $\beta$), the various conservation laws [Eqs.~(\ref{baryonnum}) and (\ref{electronnum})] and conditions for phase equilibrium [Eqs.~(\ref{muneq}),~(\ref{mupeq}),~(\ref{betaQ}),~(\ref{weakQ}),~(\ref{mecheq})] must be applied. 
The solution of these equations converts the $y_i$ and $f$ from independent variables to functions of $\nb$ and $\eta$. Thus, the state variables also become functions of $\nb$ and $\eta$ according to the rule
\begin{eqnarray}
Q(\nb,y_i,y_j,...,\eta) &\rightarrow& Q_{\beta}[\nb,y_i(\nb,\eta),y_j(\nb,\eta),...,\eta]  
\nn \\
&=& Q_{\beta}(\nb,\eta)~. \nn 
\end{eqnarray}
Note that the upper- and lower-density boundaries of the mixed phase correspond to 
$f_{\beta}(\nb,\eta) = 0$ and 1, and depend on $\eta$.

The adiabatic speed of sound in the mixed phase is obtained by first calculating the expression
\be
\csqad(\nb,y_i,f,\eta) = \left.\frac{\p P^*}{\p \nb}\right|_{y_i,f,\eta}
\left(\left.\frac{\p \ep^*}{\p \nb}\right|_{y_i,f,\eta}\right)^{-1}
\ee
and then evaluating it for NSM
\be
c_{\rm{ad},\beta}^2(\nb,\eta) = \csqad[\nb,y_{i,\beta}(\nb,\eta),f_{\beta}(\nb,\eta),\eta]~.
\ee
On the other hand, the equilibrium sound speed is given by the total derivatives of the pressure and the energy density with respect to the baryon density \textit{after} the enforcement of NSM equilibrium,
\be
\csqeq = \frac{dP^*_{\beta}}{d\nb}\left(\frac{d\ep^*_{\beta}}{d\nb}\right)^{-1}~.
\ee
%

\section{Nonradial neutron star oscillations }
Neutron stars are expected to oscillate in many modes corresponding to different restoring forces. Pressure-supported modes including $f$-(fundamental) and $p$-(pressure) modes are sensitive to stellar structure. The $f$-mode frequency approximately scales with the mean density and is universally correlated with the tidal deformability and the moment of inertia~\cite{Andersson:1997rn,Lau:2009bu,Zhao:2022tcw}. $p$-mode oscillations are more confined toward the surface of the NS and are thus sensitive to the EOS at lower density~\cite{Kunjipurayil:2022zah}. Both $f$- and $p$-modes are sensitive to the bulk pressure and not sensitive to detailed chemical composition. We have verified that the novel construction of first-order phase transitions in this work  does not play a significant role due to the universal relation between the oscillation frequencies and other NS observables. \\

In this paper, we study the $g$-mode, the fluid mode with gravity as the restoring force. The $g$-mode oscillation acquires nonzero frequency because there is a gradient of chemical composition or a first-order phase transition between the two phases \cite{RG92,Finn:1987}. A universal relation between the chemical $g$-mode frequency and lepton fraction was discovered recently~\cite{Zhao:2022toc} providing key information about the nuclear symmetry energy at high density. A $g$-mode due to a density discontinuity from a phase transition can be understood as a special version of a $g$-mode due to chemical composition changes, since matter on the low-density side can be treated as having a different composition from that on the high-density side. This situation occurs when matter does not instantaneously change phase upon passing through the phase transition boundary~\cite{Pereira:2017rmp}. The discontinuity $g$-mode is most sensitive to the local gravity and the density discontinuity at phase transition~\cite{Miniutti:2002bh,Zhao:2022tcw}. \\

At high temperature relevant to neutron star mergers, the compositional $g$-mode can be suppressed~\cite{Lozano:2022qsm}. However, another branch of $g$-mode can also have nonzero frequency when adiabatic compression of the NS matter is not in thermal equilibrium with the matter in hydrodynamic equilibrium~\cite{McD83}. These are very-low-frequency modes because thermal pressure is negligible in the cores of neutron stars when temperature $T\lesssim 10^7$ K~\cite{McD83}. For $T\gtrsim 10^{10}$ K, the thermal $g$-mode has comparable frequency to the compositional $g$-mode~\cite{Kuan:2022bhu}. Recent core-collapse supernova simulations suggest that the thermal $g$-mode could dominate when there is a large entropy gradient~\cite{Jakobus:2023fru}. \\

In this work, we consider only the zero-temperature EOS for hybrid NSs with the novel framework of a first-order transition.
We focus on the lowest order nonradial $g$-mode oscillation ($\ell=2$) arising from a gradient in the chemical composition. This oscillation mode couples directly to gravitational waves and has a frequency of a few hundred Hz for NSs which lies in the band of gravitational wave observations~\cite{Tran:2022dva,Kumar:2021hzo}. Assuming the chemical composition does not change in a period of oscillation, the local $g$-mode frequency $\nu_g$ is determined by the \bv~frequency,
\begin{eqnarray}
\nu_g^2&=& g^2\left(\frac{1}{c_{eq}^2}-\frac{1}{c_{ad}^2}\right) e^{\nu-\lambda}\,, \label{eq:BV_frequency}
\end{eqnarray}
where $\nu$ and $\lambda$ are the temporal and radial metric functions. The \bv~frequency depends on density and chemical composition that vary across the NS. We show in Sec.~V C the difference between the inverse squared sound  speeds and the bracket on the right-hand side of Eq. (\ref{eq:BV_frequency}) for the various models studied. With the correct boundary condition and perturbation fluid equations, one can find global oscillation modes, known as $g$-modes driven by local buoyancy oscillations. 
{Such $g$-modes have been studied for hybrid NSs with Gibbs construction and under the Cowling approximation \cite{Jaikumar:2021jbw,Kumar:2021hzo}. In this work, we solve the $g$-mode with linearized theory of full general relativity.} Detailed methods to calculate the $g$-modes with and without the Cowling approximation can be found in our previous work~\cite{Zhao:2022toc}. 

\section{Results}\label{sec:result}
In this section we demonstrate the effect of changing the local-to-total lepton ratio on the EOS and its composition, associated structural and tidal properties of NSs, the two sound speeds, and the resulting \gm~frequencies. We also show plots pertaining to the EOS and particle fractions of a crossover application. All results refer to neutron star ($\beta$-equilibrated) matter. 
\subsection{Equation of state}
\begin{figure}[htbp!]
    \includegraphics[width=\linewidth]{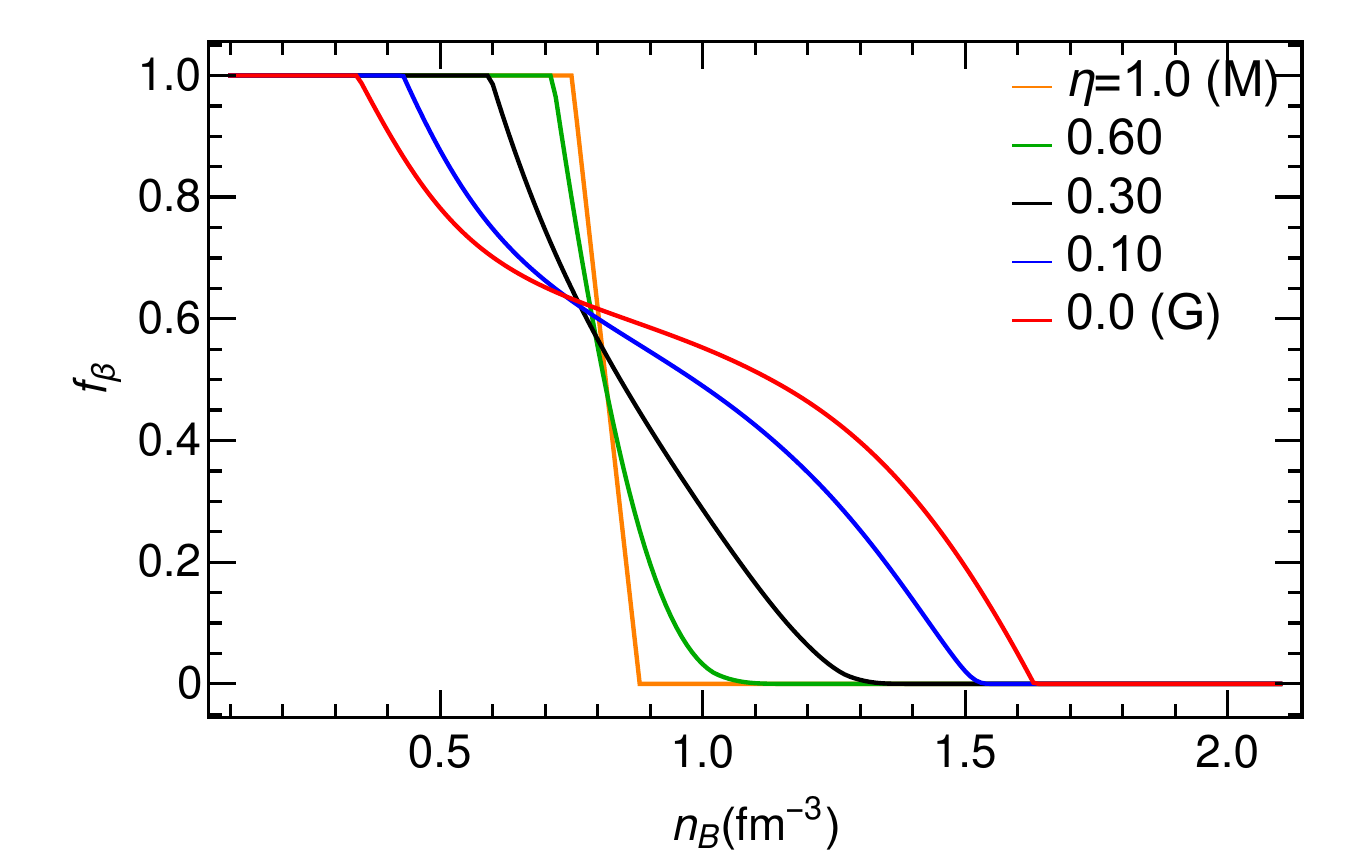}
    \caption{Nucleon-to-baryon fraction vs baryon density for the indicated values of the local-to-total lepton ratio, $\eta$.}
    \label{fig:fbeta}
\end{figure}
The change in the nucleonic content of the mixed phase is shown in Fig.~\ref{fig:fbeta} for five different implementations of charge neutrality. The decrease in $f_{\beta}$ is steeper as the Maxwell limit (of high surface tension and thus LCN) is approached; that is, the mixed phase becomes narrower in terms of density. This indicates that first-order transitions with sharper phase separation (Maxwell-like, ``stiff") undergo a faster compositional change that can impact the $g$-mode frequency more severely than transitions where extensive phase mixing occurs (Gibbs-like, ``soft"). The approximately common intersection point of the various curves occurs at, roughly, $f_{\beta}=2/3$ near the density $n_t$ at which the energy densities of the pure phases are equal. We note that, for the models and parametrization used herein, the boundaries of the Gibbs and the Maxwell mixed phases are (0.34, 1.63) and (0.75, 0.88)~fm$^{-3}$, respectively. 

\begin{figure}[htbp!]
    \includegraphics[width=\linewidth]{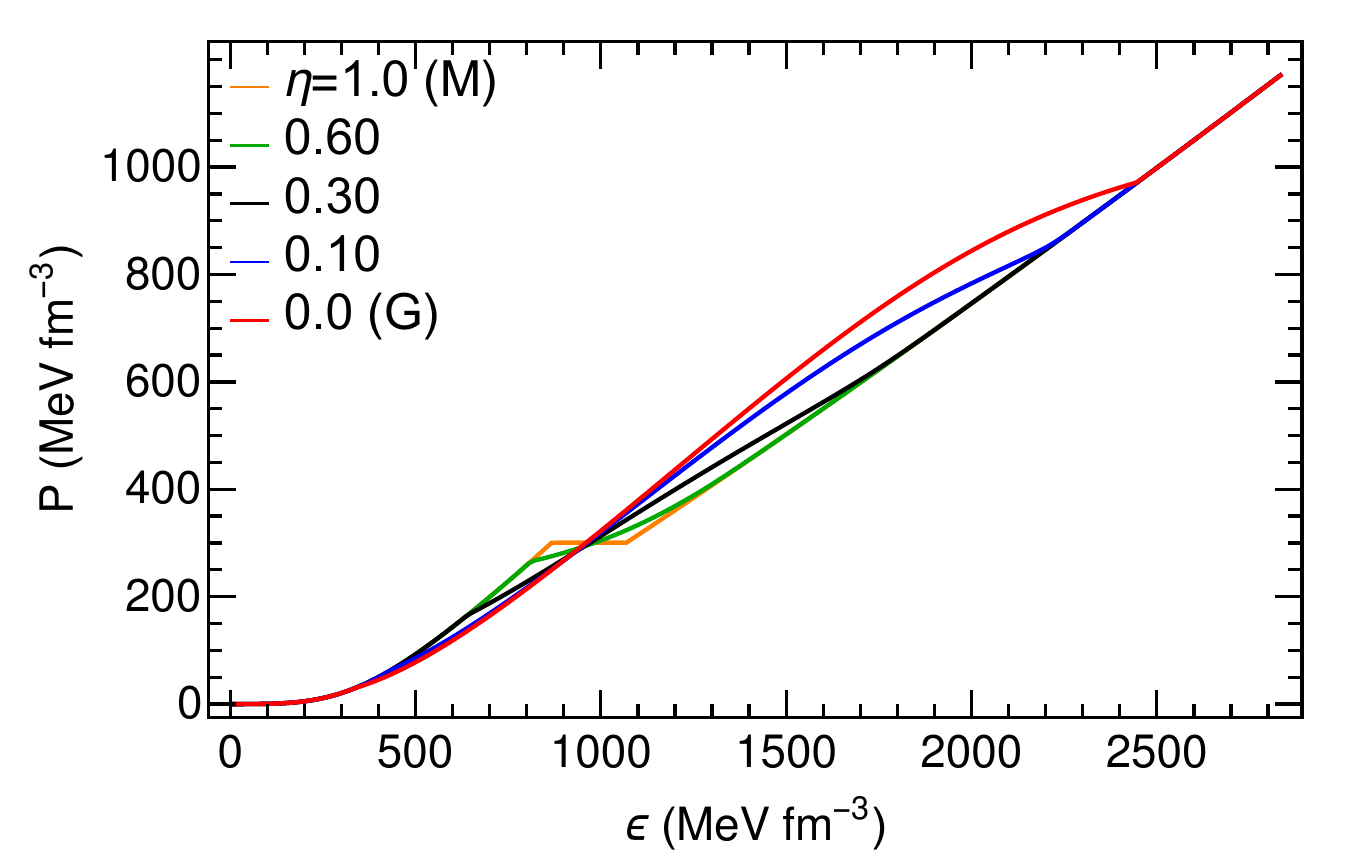}
    \caption{A representation of the EOS of $\beta$-equilibrated matter in the pressure vs energy density plane for various $\eta$'s.} 
    \label{fig:eos}
\end{figure}

Figure~\ref{fig:eos} is a representation of the effect of varying $\eta$ on the EOS in the pressure vs energy density plane for $NQe\mu$ matter. It follows the trends already seen in Fig.~\ref{fig:fbeta} of more Maxwell-like behavior with increasing $\eta$ and a correspondingly smaller mixed phase. Our EOS includes a wide variety of possibilities between the Maxwell and Gibbs constructions, some of which are very similar to EOSs of 
the quark-hadron phase calculated using the Wigner-Seitz approximation (WSA); see, e.g.,~\cite{Maruyama:2007ey,Yasutake:2014oxa,Wu:2018zoe,Maslov:2018ghi}. 
Thus, we may interpret the present framework as one that recasts the complicated Coulomb and surface problem of the WSA into an easier form involving only lepton phase space, with local leptons increasing the energy of the system mimicking the effect of surface energy. \\

\begin{figure}[htbp!]
    \includegraphics[width=\linewidth]{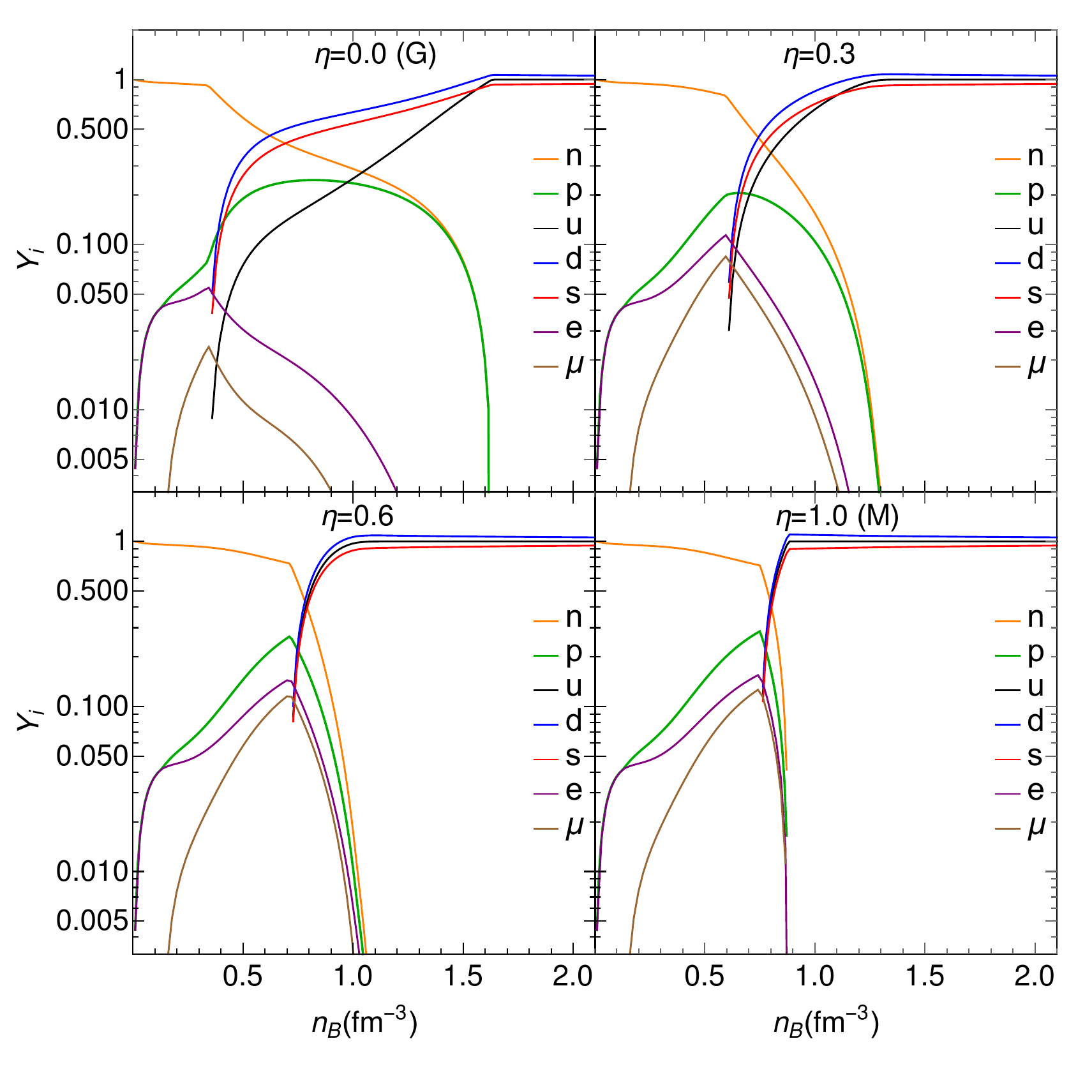}
    \caption{Various particle fractions vs baryon density for the indicated $\eta$'s.  Muons drop out of the system at intermediate to high densities (when 
    $\mu_e = m_{\mu}$ is met). Electrons are always present, but, at higher densities,  at 2 orders of magnitude less than what is shown here.}
    \label{fig:yi}
\end{figure}

Figure~\ref{fig:yi} shows the particle fractions corresponding to four different $\eta$'s: $\eta=0$ (Gibbs), $\eta=1$ (Maxwell), and two intermediate cases of $\eta=0.3$ and $\eta=0.6$. Two features are of particular interest here: (1) the total lepton fraction $y_L = y_e + y_{\mu}$ tends to the proton fraction $y_p$ with increasing $\eta$; that is, for more Maxwell-like transitions, charge neutrality for the nucleonic sector in the mixed phase is largely achieved via negatively charged leptons, whereas (2) for more Gibbs-like transitions, the negatively charged quarks $d$ and $s$ are the main counterparts to the proton. As a result, $u$ quarks are  suppressed relative to the pure-quark-phase abundances of the three species for small $\eta$'s.  \\

\begin{figure}[htbp!]
    \includegraphics[width=\linewidth]{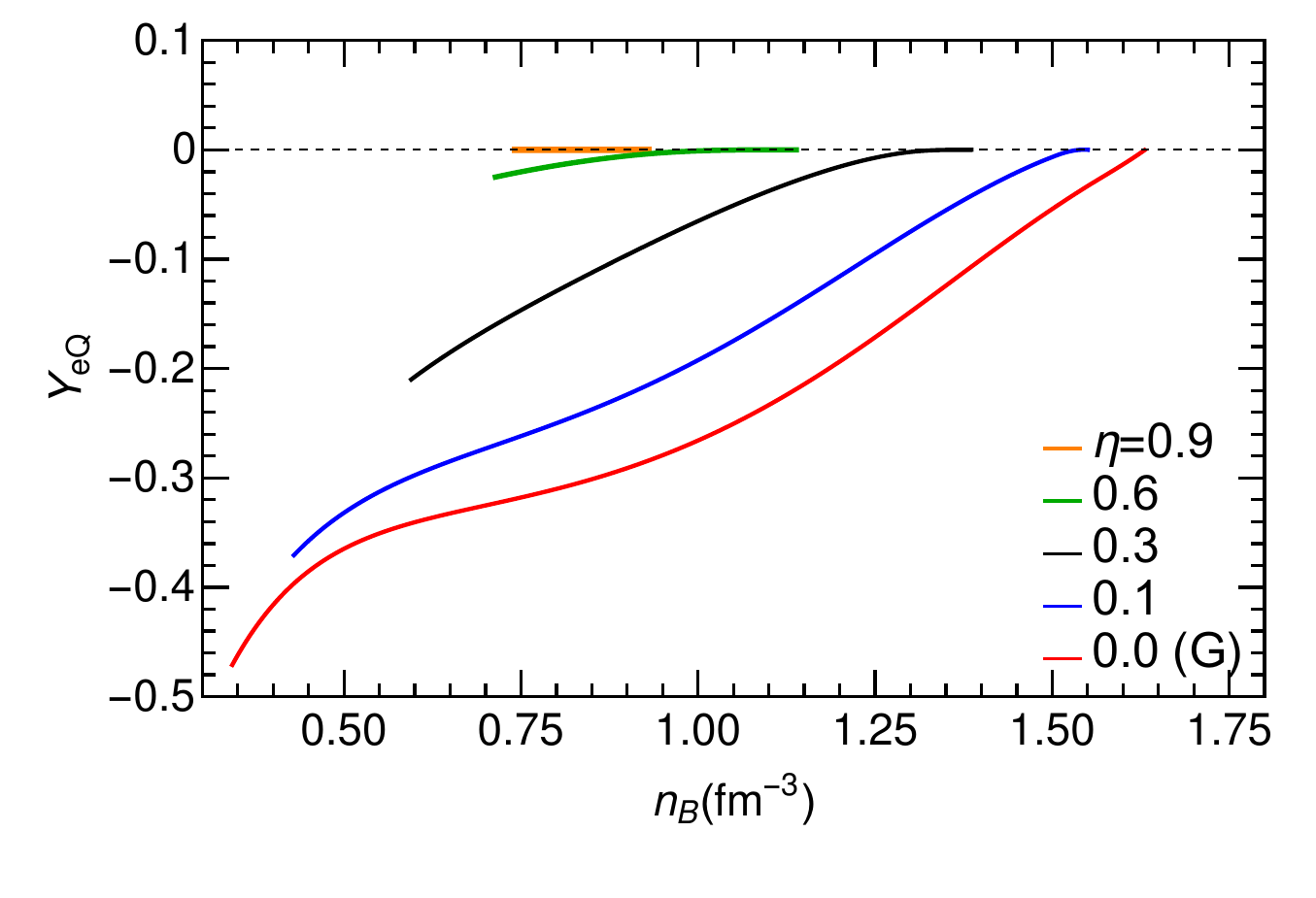}
    \caption{Fraction of electrons ensuring LCN for quarks in the mixed phase. Note the negative sign indicating opposite electric charge (positive).}
    \label{fig:yeq}
\end{figure}
This is also reflected in the fractions of the ``quark-attached" leptons that, in the mixed phase, operate as if they are positively charged. See Fig.~\ref{fig:yeq} for $y_{eQ}$; the muonic case (not shown) is qualitatively similar. Note that these fractions must be weighed by an overall factor of $(1-f) \eta$ in the calculation of the net lepton fractions. We can understand this behavior as follows: An inverse beta reaction such as $u^+ \rightarrow d^- + l^+ + \nu_l$ cannot occur in vacuum because it is endothermic. In medium, however, it can proceed by borrowing the missing energy from the system and, in doing so, lowering the latter's total energy (as desired). Clearly, positrons and antimuons, if seen in isolation, make positive contributions to the total energy density of the system. However, their presence lowers the net electron and muon fractions $y_e$ and $y_{\mu}$ (which are the physical quantities) leading to a composition with an overall lower energy (relative to the case without antileptons). It should be mentioned here that the lowest energy configuration is the one with $\eta=0$, i.e. the Gibbs case. Unsurprisingly, it is also the configuration with the lowest (largest absolute value) $y_{eQ}$ since the presence of positrons removes energy from the system---even though, when $\eta=0$, these positrons do not contribute to the net electron fraction.
\subsection{Neutron stars}
For the calculations in this section, in addition to the core EOS described previously,
we use SLy4 crust EOS for $\nb<0.05$ fm$^{-3}$~\cite{Chabanat:1997un,Douchin:2001sv}.

\begin{figure}[htbp!]
    \includegraphics[width=\linewidth]{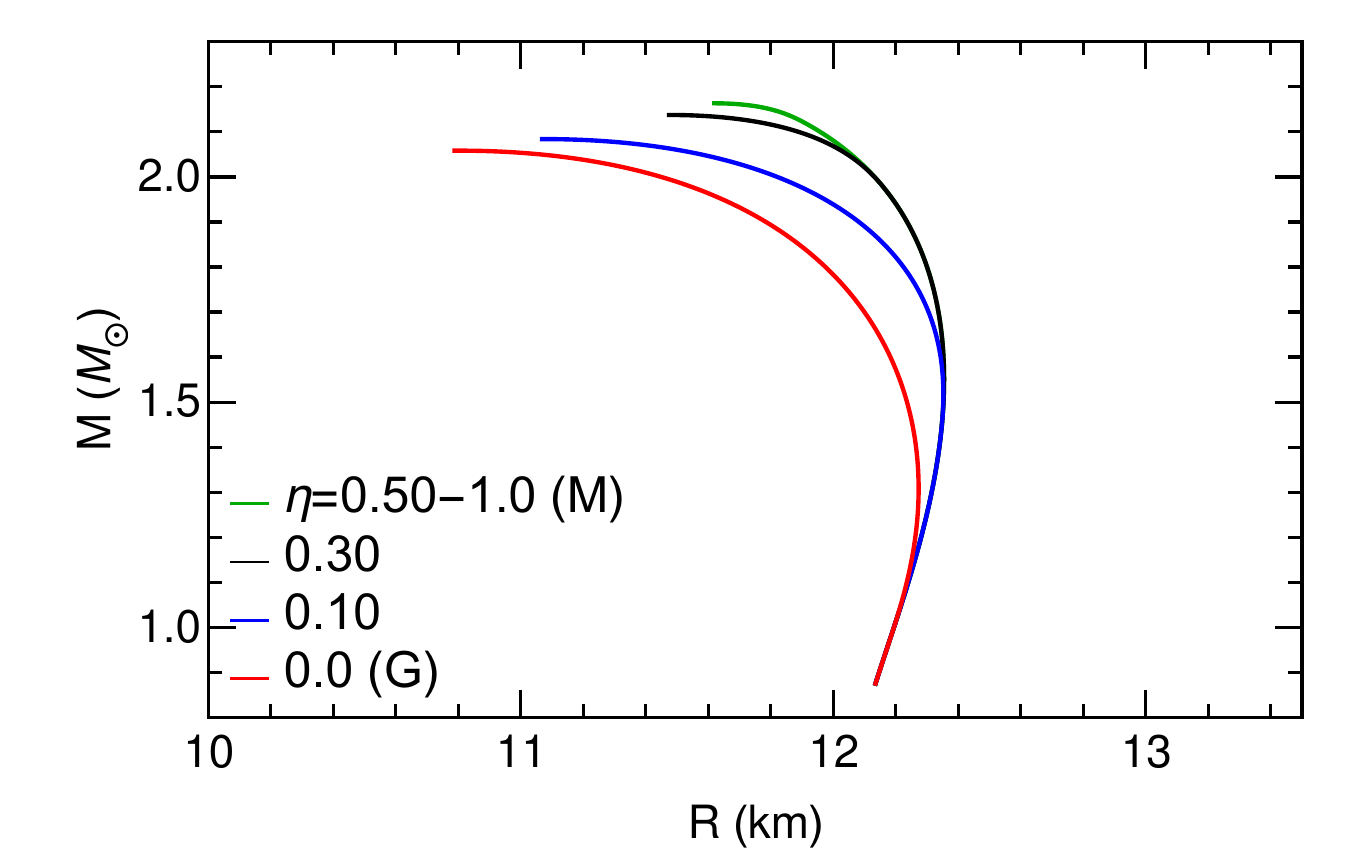}
    \caption{Neutron star mass-radius curves for various $\eta$'s. Radii and masses both move to higher values with increasing $\eta$, before leveling out for $\eta\ge 0.5$.}
    \label{fig:ns_mr}
\end{figure}
Figure~\ref{fig:ns_mr} depicts the NS mass-radius ($M$-$R$) diagram for a few chosen values of $\eta$. The maximum mass as well as the corresponding radius rise with $\eta$ from $(M,R) = (2.05 \,\Msolar, 10.8~\mbox{km})$ for $\eta=0$ (Gibbs) to $(2.17 \,\Msolar, 11.7~\mbox{km})$ for $\eta=1$ (Maxwell). The latter set is shared by all stars with $\eta\ge 0.5$ whereas a larger spread occurs as $\eta$ moves to lower values. On the other hand, stars close to the canonical mass of $1.4~\Msolar$ deviate from the general trend only for small $\eta\le 0.1$.

\begin{figure}[htbp!]
    \includegraphics[width=\linewidth]{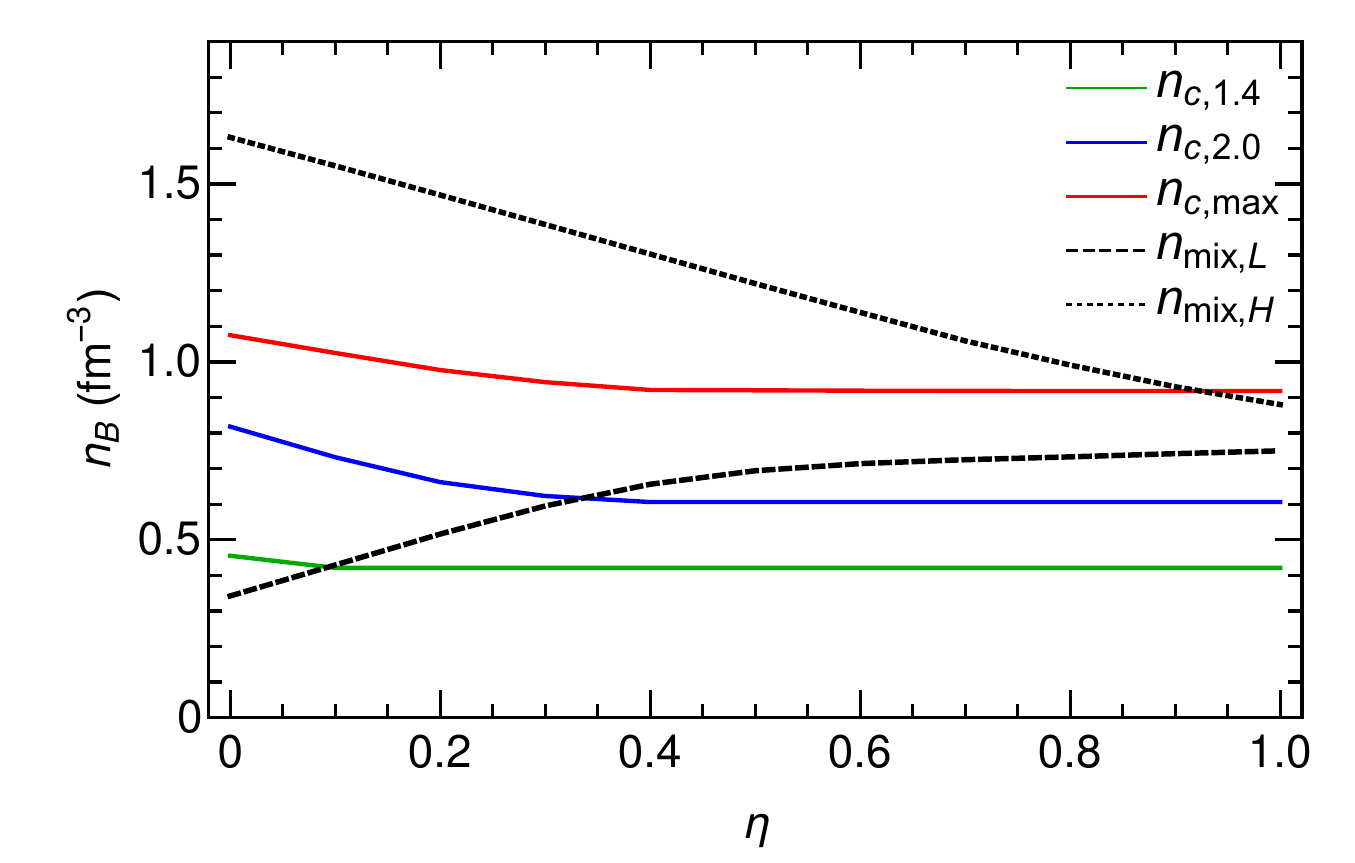}
    \caption{NS central densities for $M=1.4 \,\Msolar$ (green), $M=2.0 \,\Msolar$ (blue), and $M=\Mmax$ (red) as functions of $\eta$. The lower (L) and upper (H) density boundaries of the mixed phase are represented by the black dashed and dotted curves, respectively.}
    \label{fig:nsden}
\end{figure}
\begin{figure}[htbp!]
    \includegraphics[width=\linewidth]{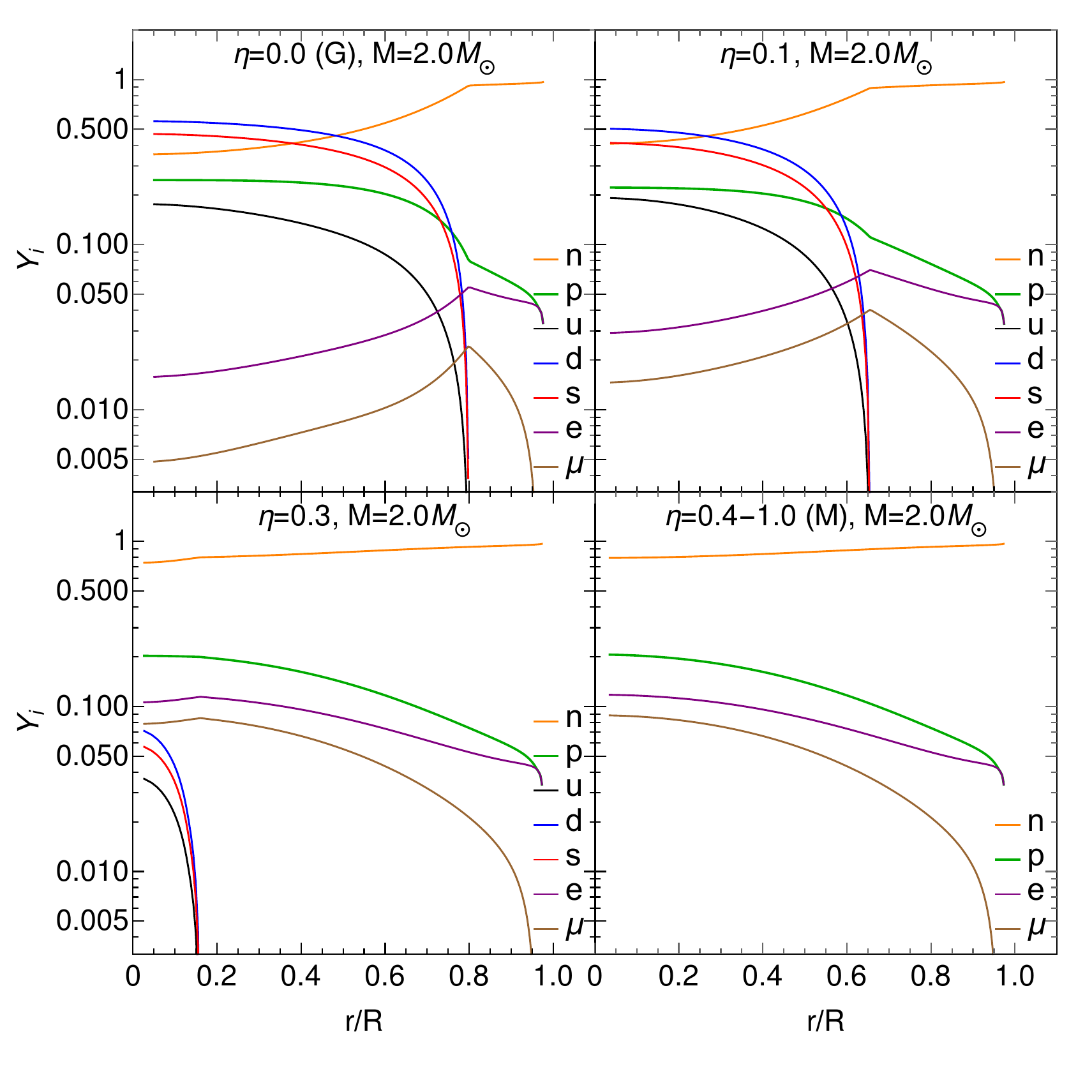}
    \caption{Various particle fractions vs NS radius for a 2.0~$\Msolar$ star. With increasing $\eta$, the quark content of the core decreases. }
    \label{fig:ns_yir20}
\end{figure}

\begin{figure}[htbp!]
    \includegraphics[width=\linewidth]{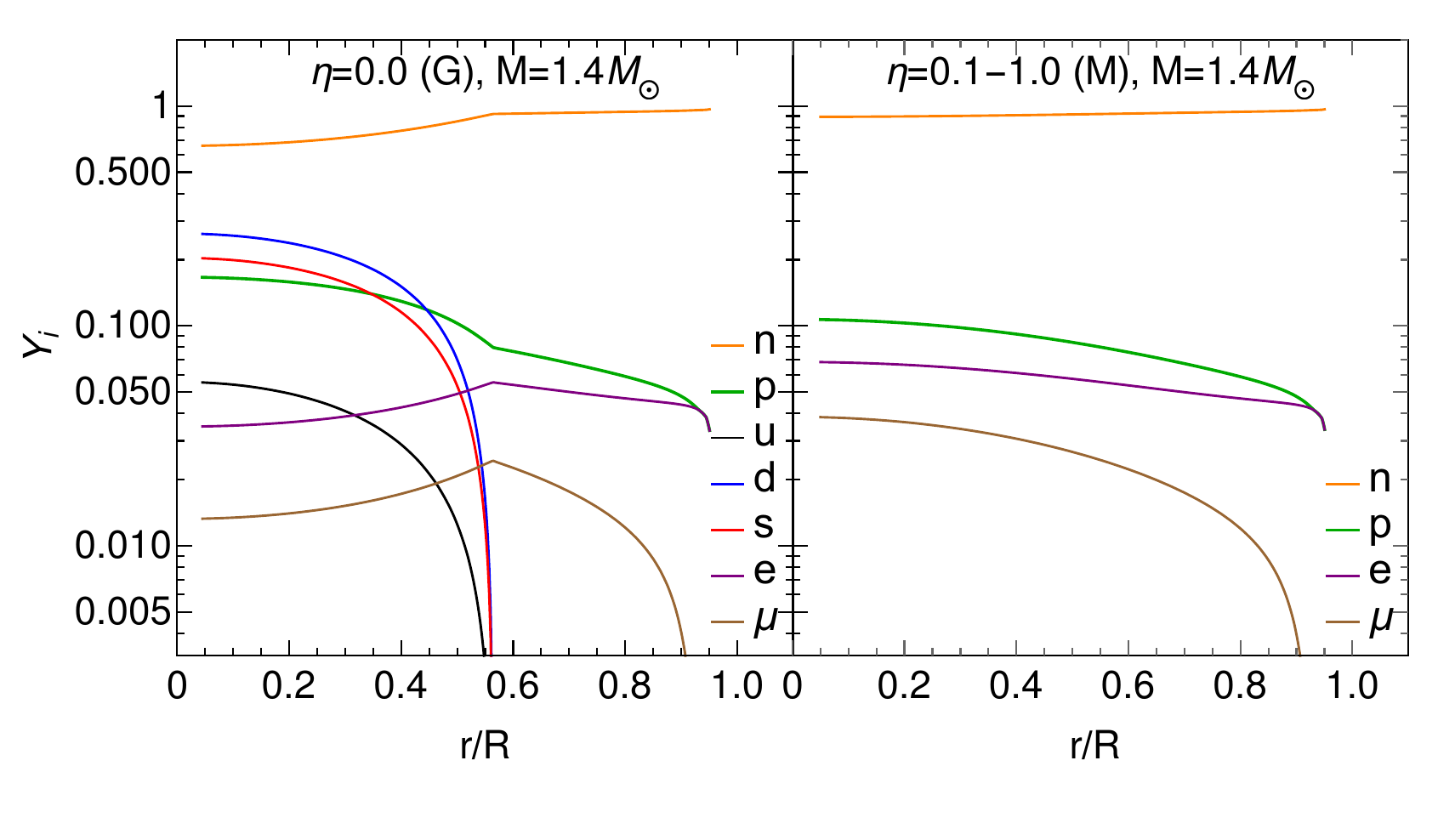}
    \caption{Various particle fractions vs NS radius for a 1.4~$\Msolar$ star. Hybrid stars of this mass can only be produced by EOSs with very soft hadron-to-quark transitions $(\eta\le0.1)$. }
    \label{fig:ns_yir14}
\end{figure}
This behavior can be better understood by turning to Fig.~\ref{fig:nsden} where the central densities of the $1.4 \,\Msolar$, $2.0 \,\Msolar$, and $\Mmax$ stars are plotted as functions of $\eta$ together with the boundaries of the mixed phase. We find that $1.4 \,\Msolar$ NSs with $\eta\ge 0.1$ are purely nucleonic, whereas $2.0 \,\Msolar$ NSs contain no quark admixture for $\eta\ge 0.35$---see also Figs.~\ref{fig:ns_yir20} and \ref{fig:ns_yir14} for the particle fractions as functions of the NS radius for these two NS masses. In the neighborhood of the maximum mass however, stars are always hybrid; the exception being $\eta\ge 0.9$, where an inner core of pure quark matter forms. This only affects the upper $0.2\%$ of the mass range. \\

\begin{figure}[htbp!]
    \includegraphics[width=\linewidth]{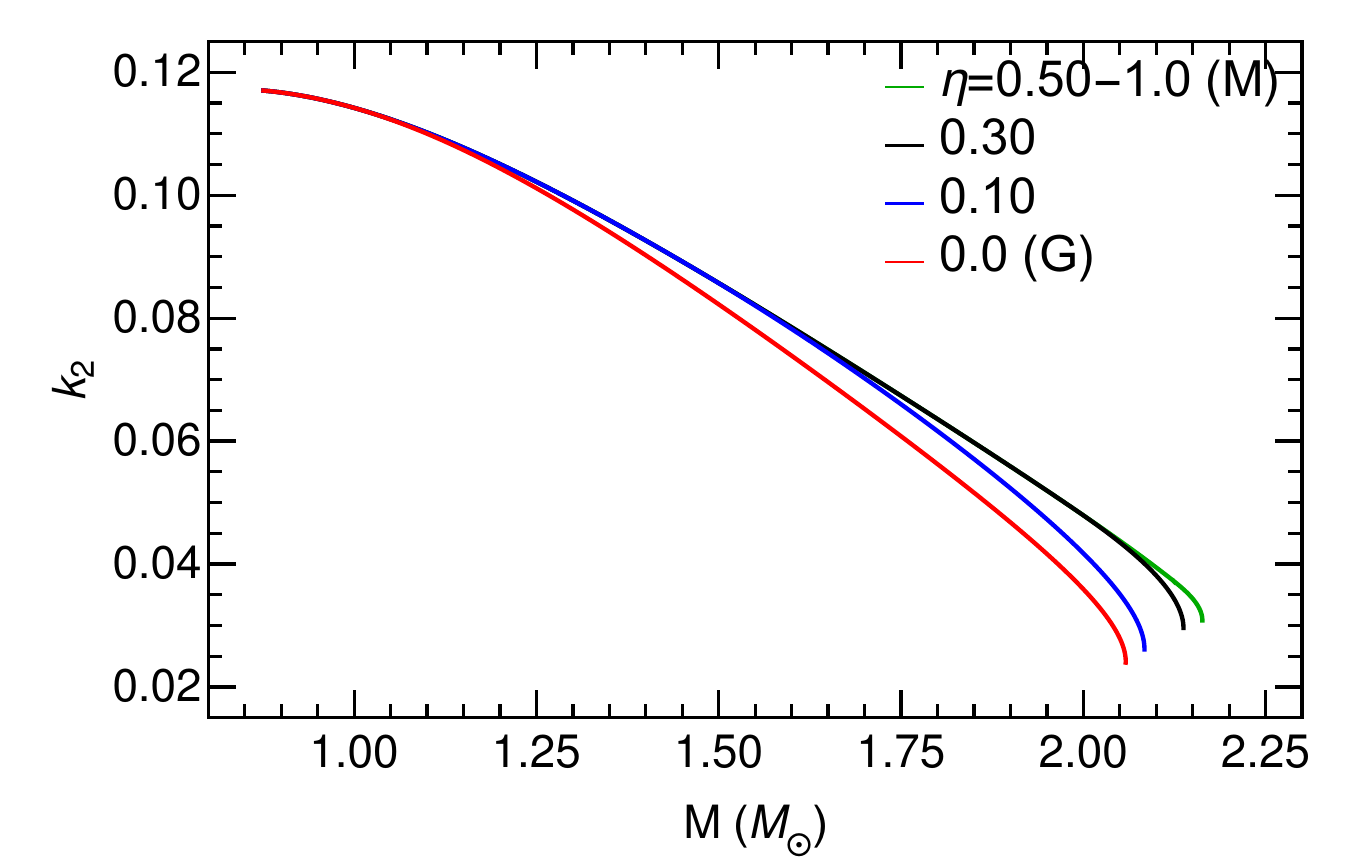}
    \caption{Love number vs neutron star mass for various $\eta$'s.}
    \label{fig:ns_k2}
\end{figure}

\begin{figure}[htbp!]
    \includegraphics[width=\linewidth]{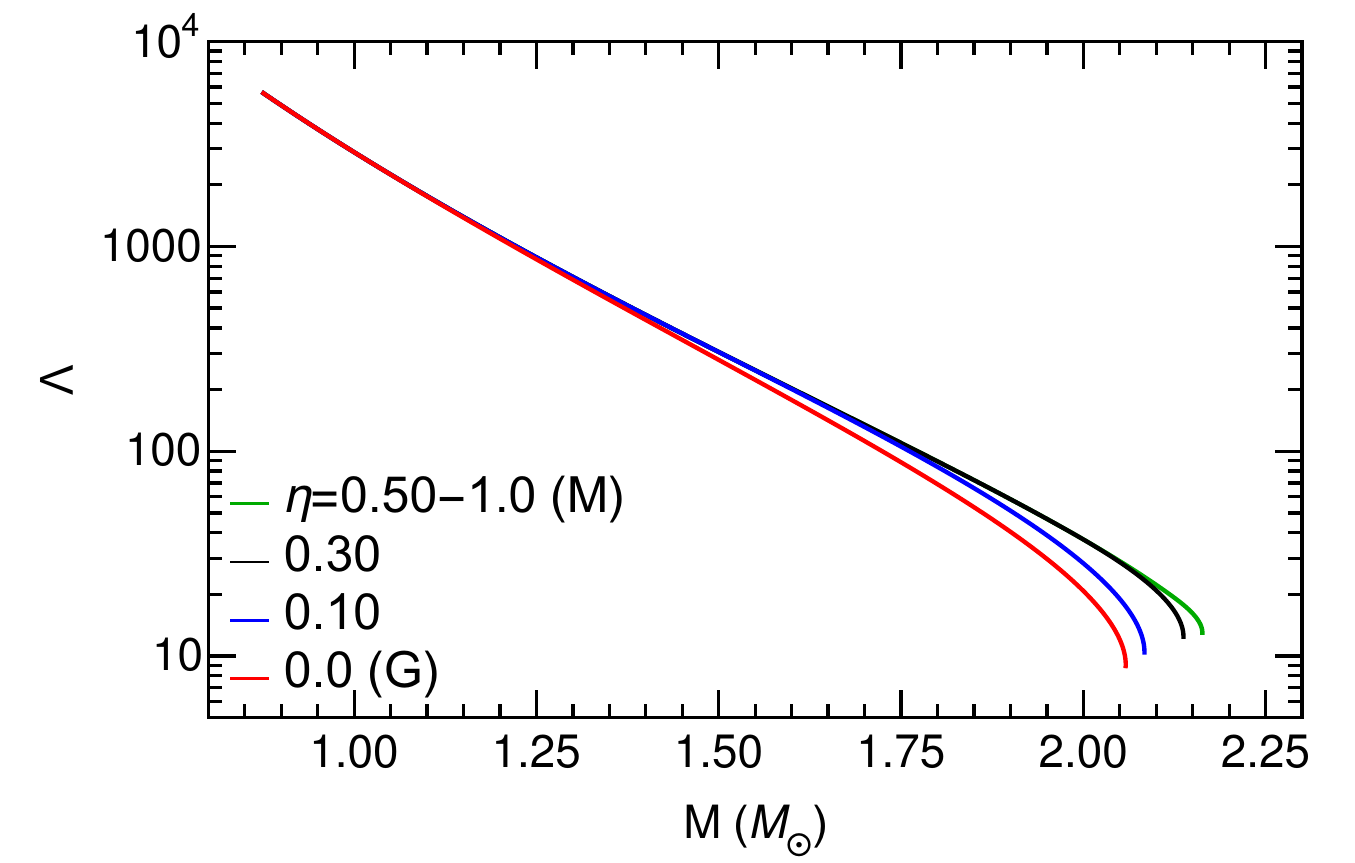}
    \caption{Tidal deformability vs neutron star mass for various $\eta$'s.}
    \label{fig:ns_lambda}
\end{figure}
The tidal properties of neutron stars such as the tidal Love number $k_2$ and the tidal deformability $\Lambda$ (Figs.~\ref{fig:ns_k2} and \ref{fig:ns_lambda}, respectively) exhibit similar trends as the $M$-$R$ diagram. That is, different implementations of charge neutrality affect stars with larger masses more severely; in part because the associated radii can change by up to $10\%$ from $\eta=0$ to $\eta=1$.

\subsection{$g$-modes}
\begin{figure}[htbp!]
    \includegraphics[width=\linewidth]{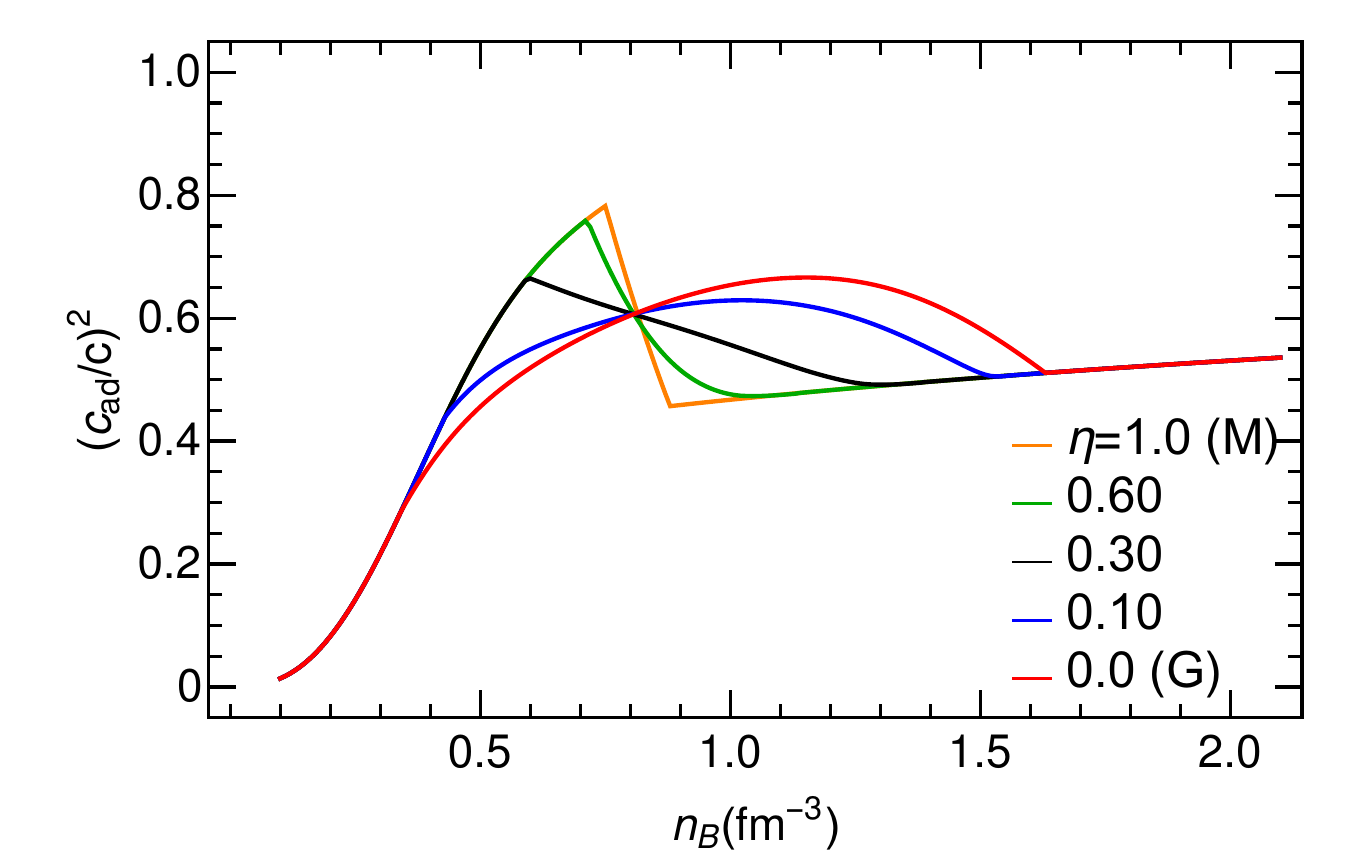}
    \caption{Squared adiabatic sound speed vs baryon density normalized to the speed of light.}
    \label{fig:cad}
\end{figure}
\begin{figure}[htbp!]
    \includegraphics[width=\linewidth]{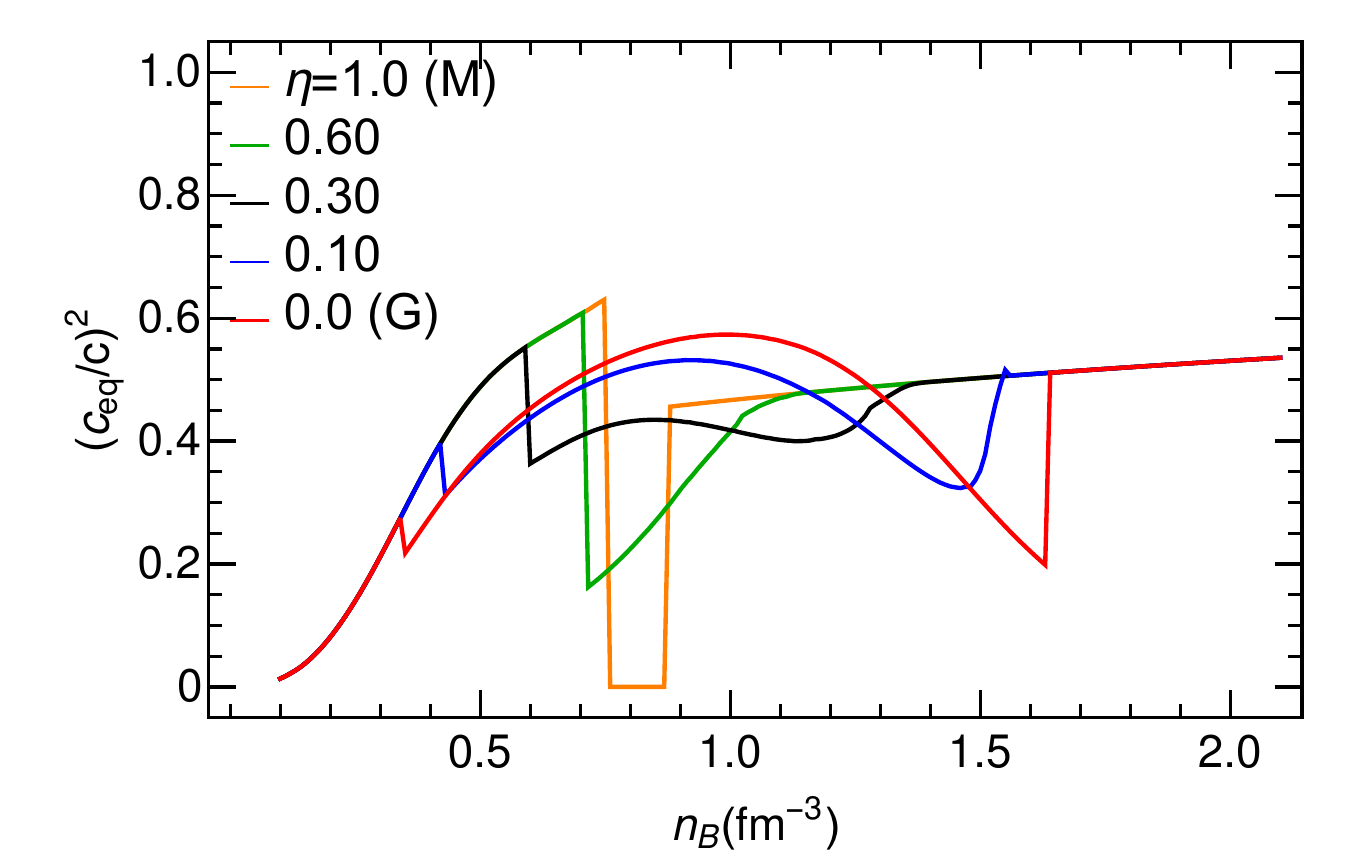}
    \caption{Squared equilibrium sound speed vs baryon density normalized to the speed of light.}
    \label{fig:ceq}
\end{figure}
\begin{figure}[htbp!]
    \includegraphics[width=\linewidth]{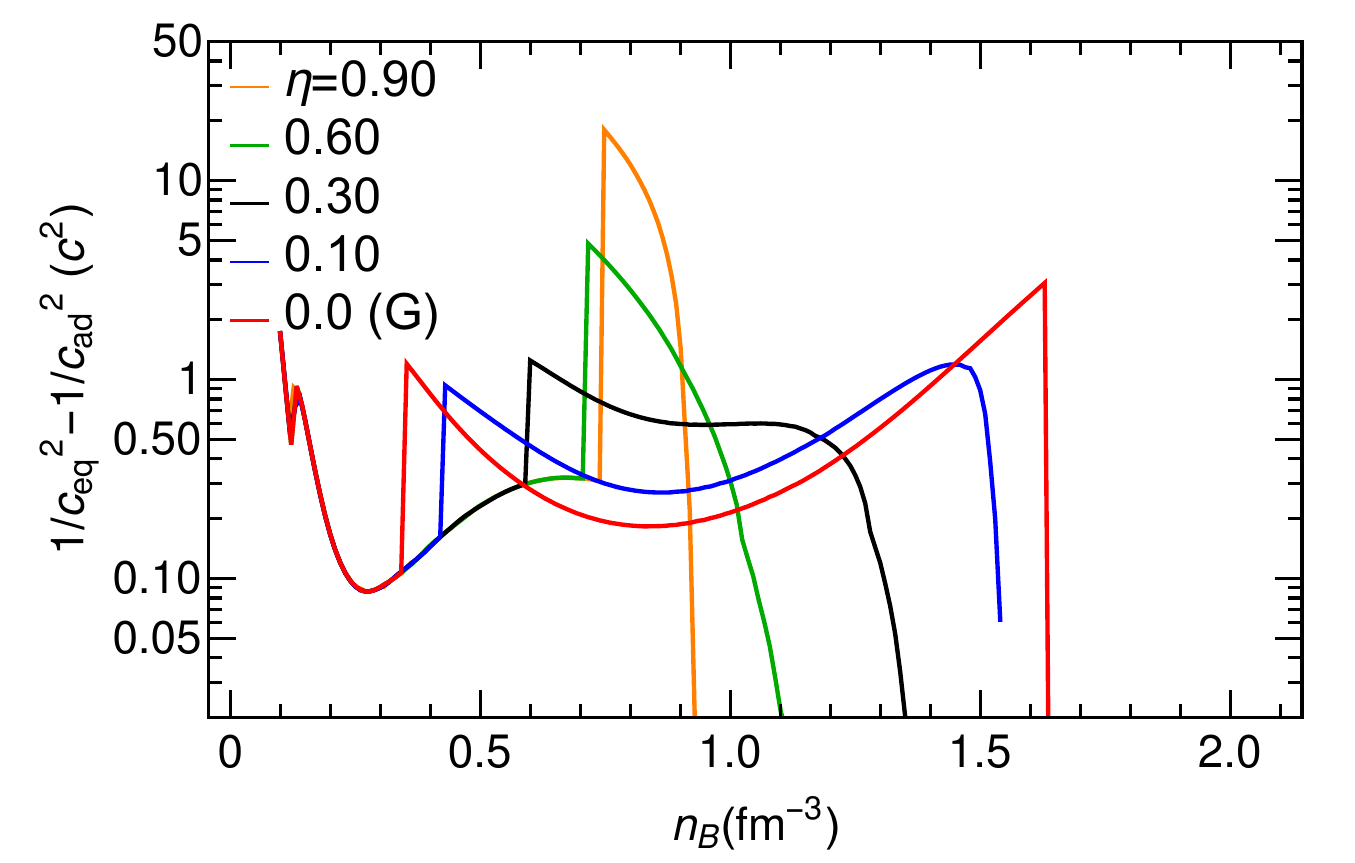}
    \caption{Difference of the inverse squared sound speeds vs baryon density. The peaks in the vicinity of $\nsat$ correspond to the appearance of muons. The peaks at intermediate densities occur at the onset of the mixed phase. Small kinks at higher densities are due to the  disappearance of muons. All curves go to (nearly) zero when the pure quark phase is reached.}
    \label{fig:delta}
\end{figure}

The adiabatic and equilibrium squared sound speeds as functions of the baryon density are shown in Figs.~\ref{fig:cad} and \ref{fig:ceq}, respectively, for five different values of $\eta$. Both increase monotonically with $\nb$ in the pure nucleonic phase, whereas the presence of both nucleonic and quark matter in the mixed phase leads to nonmonotonic behaviors. Being that the transition onset is at higher density for higher $\eta$, larger sound speeds (of either kind) are attained by the more Maxwell-like models with correspondingly sharper decreases over the mixed phase. The relative change is more pronounced in the case of the equilibrium sound speed.  Since particle fractions in pure quark matter are almost constant, see Fig.~\ref{fig:yi}, the adiabatic and equilibrium sound speeds are very nearly the same. As a result, NSs with inner cores of pure quark matter have negligible \bv~frequency as shown in Fig.~\ref{fig:delta}. 
These are the stars with $\eta\ge 0.9$ in the vicinity of the maximum-mass configuration mentioned earlier in the context of Fig.~\ref{fig:nsden}. \\

\begin{figure}[htbp!]
    \includegraphics[width=\linewidth]{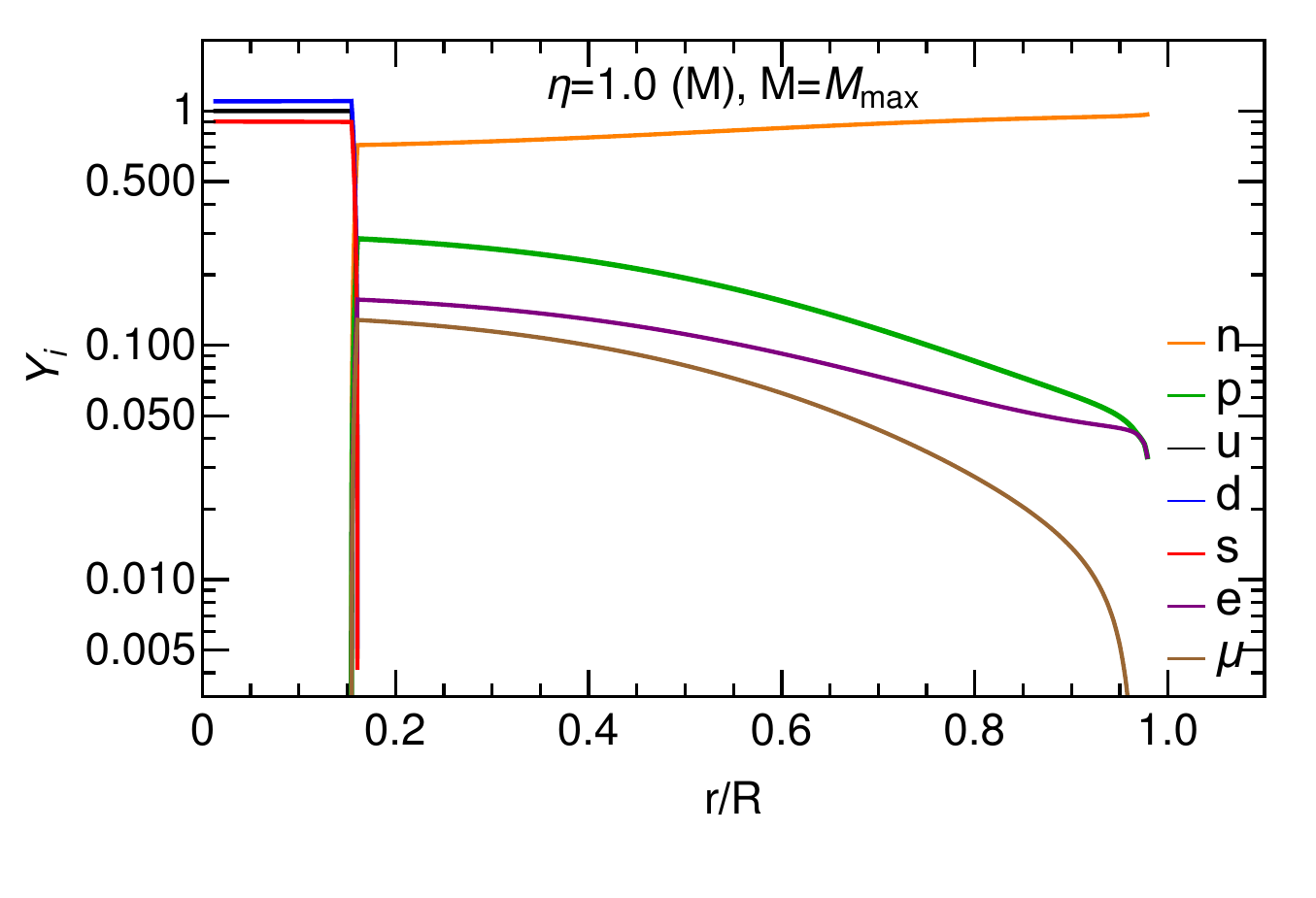}
    \caption{Particle fractions vs NS radius for the maximum-mass  star in the case of Maxwell construction. 
    The flatness of the pressure vs. density in the mixed phase leads to a density jump in the core of a neutron star from the lower boundary of the mixed phase to the upper. Consequently, the star, at any given radial distance from its center, contains either  hadronic or quark matter; not both. }
    \label{fig:mmax_yi}
\end{figure}

The influence of the growing surface tension to the difference of the inverses of the two sound speeds (Fig.~\ref{fig:delta}) is quite dramatic. While sharp peaks at intermediate-to-high densities occur for all $\eta$'s, those associated with more Maxwell-like transitions can be orders of magnitude higher than those of softer transitions. Given that this quantity enters directly in the calculation of the \bv~frequency, correspondingly strong $g$-mode signals may be produced. 
Also worth noting is that the kinks, evident on the $\eta=0.3$ and $\eta=0.6$ curves at $\nb \simeq1.3$ and 1.0 fm$^{-3}$, respectively, occur when muons exit the system. For the Maxwell construction case with $\eta=1.0$, the particle fractions are shown in Fig. \ref{fig:mmax_yi} \\
\begin{figure}
    \includegraphics[width=\linewidth]{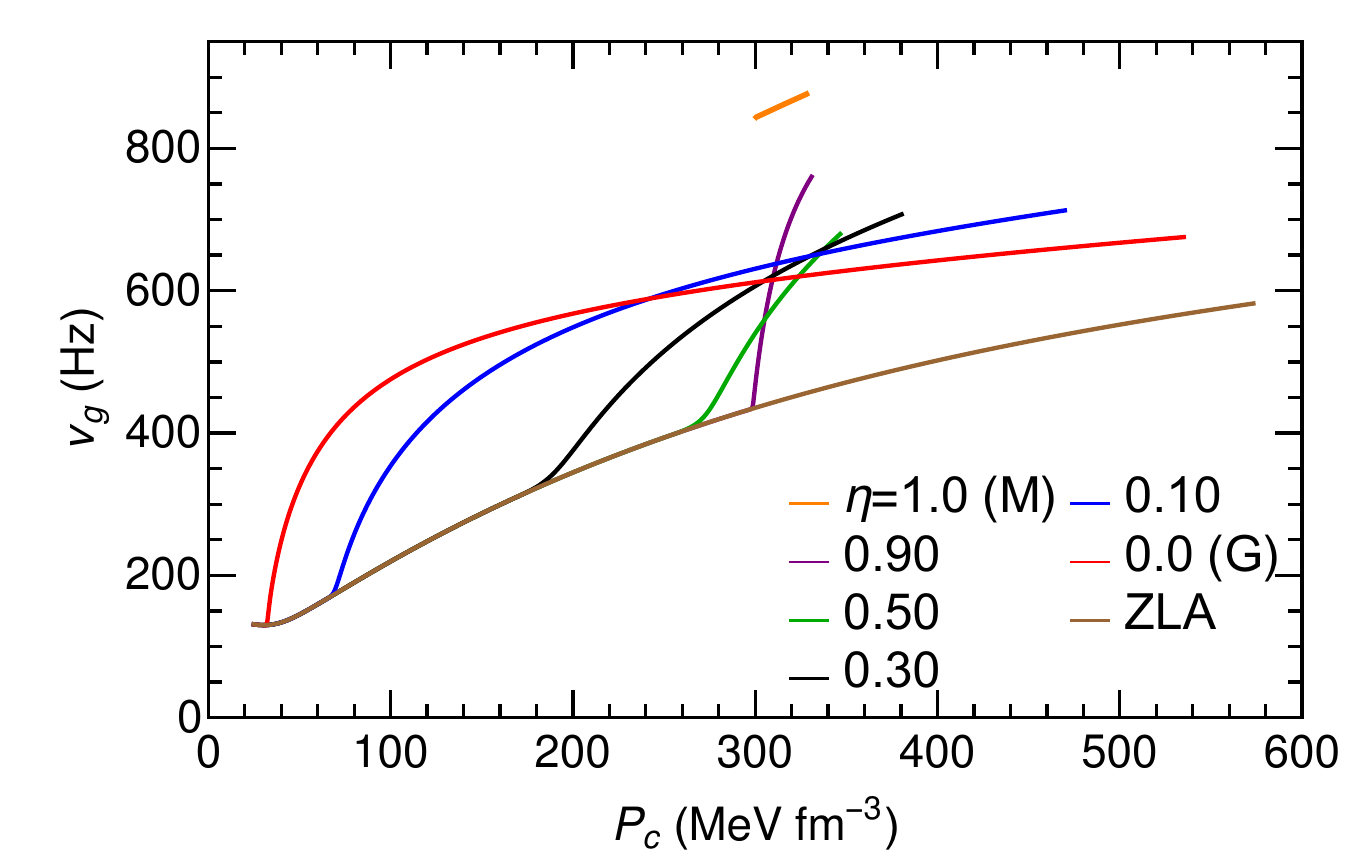}
    \caption{The $g$-mode frequency vs NS central pressure. The small orange line segment at $(P_c,\, \nu_g)\simeq (300~\mbox{MeV~fm}^{-3},\,850~\mbox{Hz})$ corresponds to the discontinuity $g$-mode generated by the Maxwell construction. }
    \label{fig:frequency_pc}
\end{figure}
\begin{figure}
    \includegraphics[width=\linewidth]{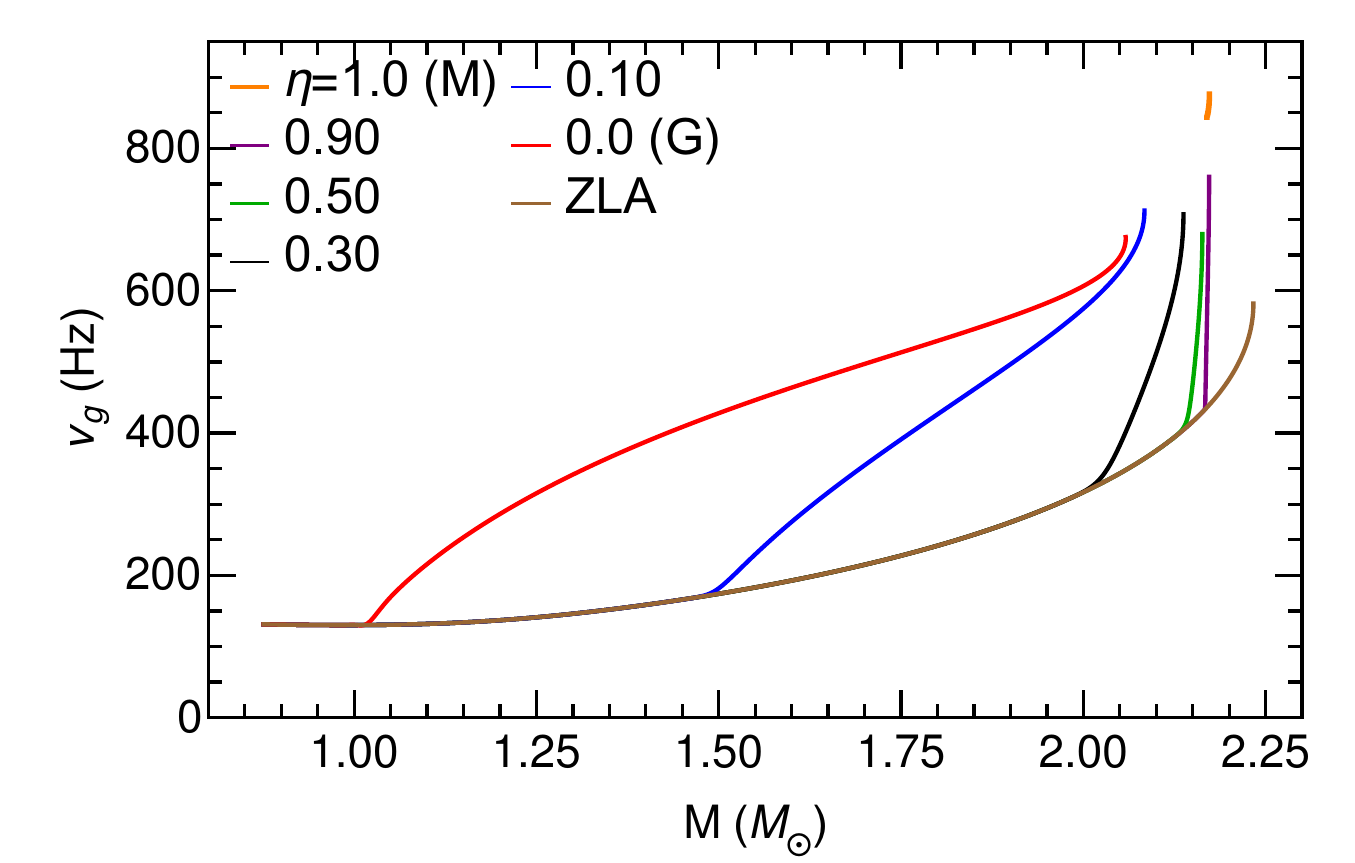}
    \caption{The $g$-mode frequency as a function of NS mass. The rapid rise occurs at progressively higher densities with larger $\eta$ and is increasingly steeper. Note the small orange line segment at $(M,\,\nu_g)\simeq (2.2~\Msolar,\,850~\mbox{Hz})$ corresponding to the discontinuity $g$-mode of the Maxwell construction.  }
    \label{fig:frequency_mass}
\end{figure}
As shown in Figs.~\ref{fig:frequency_pc} and \ref{fig:frequency_mass}, the $g$-mode frequency increases smoothly with mass for hadronic NSs. The $g$-mode frequency for hybrid NSs increases rapidly when quarks appear at the center of the NSs because their mixed phases have larger \bv~frequencies. Since a transition model with a larger surface tension has larger composition gradients and forms a narrower and higher peak in the \bv~frequency, the $g$-mode frequency increases more rapidly for a larger $\eta$. \\

In the case of a Maxwell transition where the quark-hadron mixture vanishes (see Fig.~\ref{fig:mmax_yi}), the chemical $g$-mode frequency is reduced to a discontinuity $g$-mode, which is discontinuous from the hadronic (ZLA) branch; see Figs.~\ref{fig:frequency_pc} and \ref{fig:frequency_mass}. The discontinuity $g$-mode frequency has been widely studied in the slow-conversion limit~\cite{Finn:1987,Miniutti:2002bh,Tonetto:2020bie,Zhao:2022tcw,Rodriguez:2021sgk} without involving chemical composition. At the slow-conversion limit, matter does not instantaneously change phase upon passing through the phase transition boundary. Indeed, a $g$-mode due to a density discontinuity from a phase transition can be understood as a special version of a $g$-mode due to chemical composition changes, since matter on the low-density side can be treated as having a different composition from that on the high-density side. When $\eta$ approaches 1 from below, the $g$-mode frequency goes toward the Maxwell case quickly. Here we include the adiabatic sound speed off equilibrium, and verify that the discontinuity $g$-mode frequency is not sensitive to the detailed chemical composition of hadronic or quark matter. To our knowledge, this is the first work explicitly showing that the compositional $g$-mode in a hybrid NS reduces to a discontinuity $g$-mode at the limit of Maxwell construction.
\subsection{Crossovers}
\begin{figure}[htbp!]
    \includegraphics[width=\linewidth]{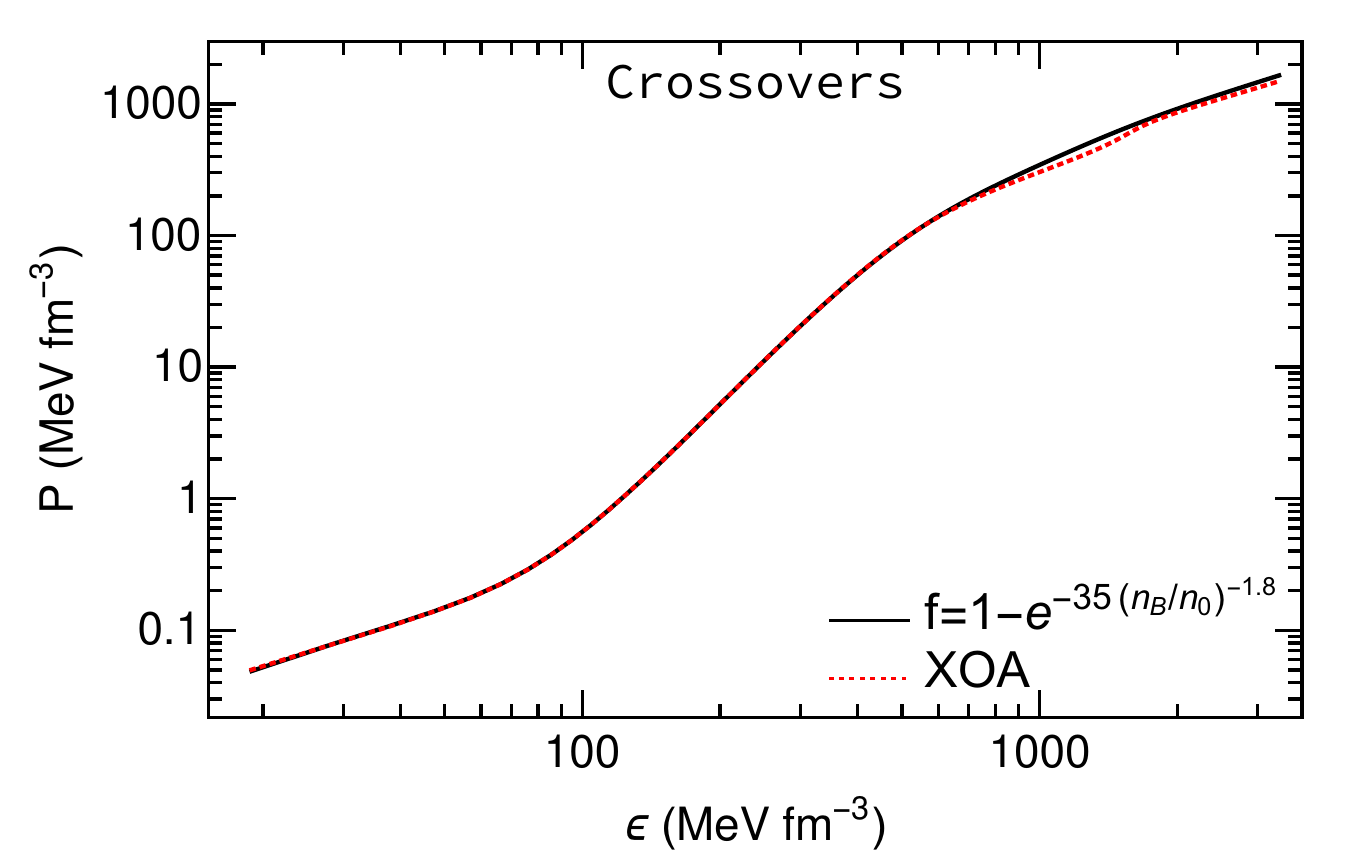}
    \caption{Comparison of the XOA parametrization of the crossover EOS of \cite{Constantinou:2021hba} based on the 
    Lattice QCD-inspired model of \cite{Kapusta:2021ney} and a crossover EOS constructed in the current scheme with $f(\nb)=1-\exp[-35~(\nb/\nsat)^{-1.8}]$.}
    \label{fig:xoeos}
\end{figure}
\begin{figure}[htbp!]
    \includegraphics[width=\linewidth]{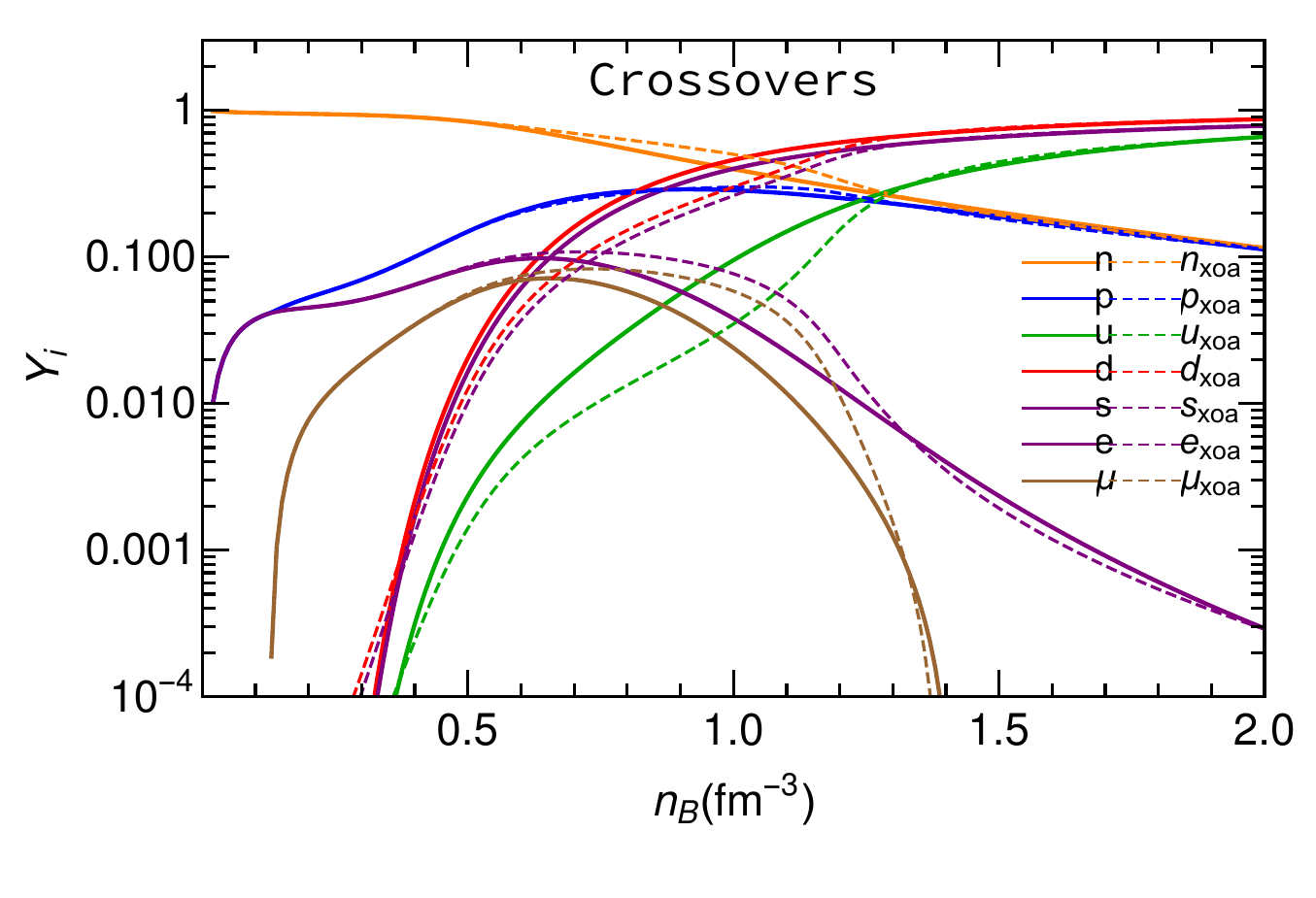}
    \caption{Comparison of the composition of $\beta$-equilibrated matter vs. baryon density for the same EOSs as in Fig.~\ref{fig:xoeos}.}
    \label{fig:xoyi}
\end{figure}
In the final two plots of this section (Figs. \ref{fig:xoeos} and \ref{fig:xoyi}), we show a comparison of the EOS and of the particle fractions corresponding to a crossover with a hadron-to-quark fraction $f(\nb)=1-\exp[-35~(\nb/\nsat)^{-1.8}]$ in the present framework and to the XOA parametrization of Ref.~\cite{Constantinou:2021hba} which provides a crossover EOS between the ZLA and vMIT equations of state as implemented in Kapusta and Welle~\cite{Kapusta:2021ney}. The various quantities are qualitatively similar even though quantitative differences exist. The use of a composition-dependent $f(\nb,y_i)$ can, presumably, improve agreement but such an undertaking is beyond the scope of this work.
%
\section{Summary and conclusions}
In this work, we have devised a thermodynamically consistent method to calculate neutron-star EOS properties when a first-order phase transition within the star lies in between the familiar Maxwell and Gibbs constructions. The implementation of this approach combines both the local and global charge neutrality conditions characteristic of the Maxwell and Gibbs constructions, respectively. Overall charge neutrality is achieved by dividing the leptons (electrons and muons) to those that take part in local charge neutrality (Maxwell) and those that maintain global charge neutrality (Gibbs). Accounting for both possibilities in conjunction with the conditions of baryon and lepton number conservation enables the calculation of the EOS upon minimizing the total energy density with respect to the various particle fractions, which generates the necessary phase-equilibrium equations. This method circumvents addressing the poorly known surface tension between the two phases microscopically (as, for example, in the calculation of the core pasta phases via the Wigner-Seitz approximation). \\

To be specific, we have considered the case of baryon (nucleons)-to-quark phase transitions. Separate model EOSs are used to describe the pure phases that contain nucleons and quarks, respectively. In the region of the phase transition, the mixed phase is characterized by the fractional volume $f$ occupied by nucleons which is solved for each baryon density using the aforementioned phase-equilibrium equations. Charge neutrality is achieved partially locally and partially globally with the aid of a new variable $\eta$, 
which lies in the range $(0,1)$ with $\eta=0$ corresponding to a Gibbs construction and $\eta=1$ to a Maxwell construction. The quantity $\eta$ serves as a proxy for the surface or interface tension between the two phases.  For a very large surface tension, the Maxwell construction with local charge neutrality is appropriate, whereas for a very small surface tension, the Gibbs construction with global charge neutrality applies. For intermediate surface tension, the boundary between the two phases is blurred and charge neutrality in the ambient phase is fulfilled both locally and globally.  \\

{The exact relation between the variable $\eta$ and the surface tension is contingent upon the particular microscopic approach used in calculating the latter. Such a relation is by no means unique being that many models for the surface tension exist in the literature \cite{Alford:2001zr,Mintz:2009ay,Palhares:2010be,Lugones:2013ema,Lugones:2018qgu,Fraga:2018cvr,Schmitt:2020tac,Ju:2021hoy}. Moreover, a density-dependent $\eta$ [that is, inclusion of Eq.~(\ref{surface}) in the equilibrium conditions] would be required for a direct comparison. } \\

{The distinguishing feature of our framework is that it enables, with a single knob $\eta \in [0,1]$, the exploration of all available EOS phase space between the Gibbs ($\eta=0$) and the Maxwell ($\eta=1$) constructions, while maintaining control over the composition. Since these constructions also correspond to the extremes of small and large surface tension (or, equivalently, complete and no phase mixing), $\eta$ can be viewed as a rough proxy for the surface tension even in the absence of a precise mapping between the two. Furthermore, it should not be difficult to parametrically tune $\eta$ such that the results of microscopic calculations are (approximately) reproduced.} \\

{The stability aspects contained in our model are as follows.  In NS matter with leptons, the pressure equation resulting from energy minimization with respect to the fractional volume occupied by hadrons $f$ assures that the pressure increases monotonically with baryon density, thus ensuring mechanical stability. Convective stability has been checked by the positivity of the \bv~frequency in Eq. (\ref{eq:BV_frequency}) throughout the star. As we discuss the zero-temperature EOS for an ideal fluid, thermal or stress related instabilities are not within the scope of discussion. } \\

{If, however, the pressures 
$P_Q$ and $P_H$ are compared, neglecting leptonic contributions, spinodal instabilities could occur; see Ref.~\cite{Constantinou:2015mna} 
for a detailed discussion in the context of a liquid-gas phase transition.   
Nucleation instability with respect to different ``pasta" phases cannot be tracked in our framework as a specific model has not been constructed as in Refs. \cite{Alford:2001zr,Mintz:2009ay,Palhares:2010be,Lugones:2013ema,Lugones:2018qgu,Fraga:2018cvr,Schmitt:2020tac,Ju:2021hoy}. 
In the approach we have developed, the precise location of charges cannot be determined. Thus a discussion of possible instabilities at the hadron-quark interface becomes impossible. } \\

Calculations of the EOSs and that of the various particle fractions for representative values of $\eta$ intermediate to $(0,1)$ are performed in this framework. The ensuing results are then utilized to calculate NS properties such the mass-radius curves, tidal deformabilities, 
adiabatic and equilibrium sound speeds, and $g$-mode oscillation frequencies.  \\

The results are in line with expectation in that the various quantities of interest transform smoothly from their Gibbs structures to those of Maxwell as $\eta$ is raised from 0 to 1. In the cases of the hadron-to-baryon fraction and of the EOS,  we find that the corresponding phase spaces between the Gibbs and the Maxwell constructions are covered in their entirety for $\eta\in [0,1]$. The composition, which favors the negatively charged quarks for the establishment of charge neutrality at small $\eta$, progressively switches over to leptons as the Maxwell limit is approached.  \\

Owing to the earlier onset of the mixed phase at smaller $\eta$'s, neutron stars with softer transitions tend to contain a higher proportion of hybrid matter at any given mass. As a result, their $M$-$R$ diagrams peak at lower values for both the mass and the associated radius; a trend which is also reflected in tidal properties such as the Love number $k_2$ and the tidal deformability $\Lambda$. However, pure-quark-matter cores are only attainable in stars with stiffer transitions because the Maxwell(-like) mixed phase covers a narrower band of densities which, for sufficiently large $\eta$ and $M$, can be exceeded by the stars' central densities. \\

A rich, nonmonotonic behavior is produced in both the equilibrium and the adiabatic sound speeds by varying $\eta$. Relative changes are more conspicuous for the Maxwell-like transitions leading to \bv~frequencies that can be orders of magnitude larger than those occurring in the opposite end of low $\eta$. \\  

The exception to the general rule of smooth change with $\eta$ is  the lowest-order $l=2$ $g$-mode frequency. It has the highest frequency among the $g$-mode family of oscillations with the fluid perturbation peaking in the core of a neutron star~\cite{Kuan:2021jmk}. Such a $g$-mode can be excited in the inspiral phase of NS mergers~\cite{Lai:1993di} causing orbital phase advance which can be measured from waveform analysis in upgraded detectors \cite{Hinderer:2016eia}. The $g$-mode frequency rises rapidly at the onset of the mixed phase, more so for stiffer transitions albeit requiring higher NS masses to be triggered. While this process advances in a regular manner for $\eta < 1$, it becomes discontinuous for $\eta=1$; that is, the $g$-mode frequency of the quark phase is discontinuous from that of the hadronic branch for a Maxwell construction. This is the first explicit demonstration of the compositional $g$-mode in a hybrid NS reducing to a discontinuity $g$-mode at the Maxwell limit. \\

Finally, we have shown how this scheme can be adapted to the description of crossovers by replacing the mechanical equilibrium condition, Eq.~(\ref{mecheq}) by a hadron-to-baryon fraction $f$ with a definite functional dependence on 
density. \\

Our results in this paper can be straightforwardly extended to finite temperature $T$ by minimizing the free energy density instead of the energy density. The conservation laws remain the same as those presented here, as do the formal expressions describing the phase equilibrium albeit with the use of finite-$T$ pressures and chemical potentials. The ensuing results will be of relevance to applications such as the short- and long-term cooling of neutron stars, simulations of binary neutron star mergers, etc.

\begin{acknowledgments}
C.C. acknowledges support from the European Union's Horizon 2020 research and innovation program under the Marie Sk\l{}odowska-Curie Grant Agreement No. 754496 (H2020-MSCA-COFUND-2016 FELLINI).
T.Z and M.P. are supported by the Department of Energy, Award No. DE-FG02-93ER40756. 
The work of S.H. was supported by startup funds from the T.D. Lee Institute and Shanghai Jiao Tong University. 

\end{acknowledgments}

\newpage

\bibliography{PRD}
\end{document}